\newcommand{\PRE}[1]{{#1}} % Use if preprint style
\documentclass[12pt,aps,prd,preprint,tightenlines,superscriptaddress,
amsmath,amssymb,%showpacs,
nofootinbib]{revtex4}
\usepackage{enumitem}
\RequirePackage[colorlinks=true
,urlcolor=blue
,anchorcolor=blue
,citecolor=blue
,filecolor=blue
,linkcolor=blue
,menucolor=blue
,pagecolor=blue
,linktocpage=true
,pdfproducer=medialab
,pdfa=true
]{hyperref}

\usepackage{amsmath,amssymb,amsthm,amsfonts}
\allowdisplaybreaks
\usepackage{slashed} 
\usepackage{graphicx}
\usepackage{subfigure}
\usepackage{dcolumn}% Align table columns on decimal point
\usepackage{hyperref}      % hypertext links %%ARXIV
\usepackage{bm}% bold math
\usepackage{epsfig}
\usepackage{epstopdf}
\usepackage{setspace}
\usepackage[usenames, dvipsnames]{color}
\usepackage{slashed}
\usepackage{comment}
\newcommand{\be}{\begin{equation}\begin{aligned}}
\newcommand{\ee}{\end{aligned}\end{equation}}
\newcommand{\beq}{\begin{equation}}
\newcommand{\eeq}{\end{equation}}
\newcommand{\beqa}{\begin{eqnarray}}
\newcommand{\eeqa}{\end{eqnarray}}

\newcommand{\ifb}{\text{fb}^{-1}}
\newcommand{\iab}{\text{ab}^{-1}}

\newcommand{\mev}{\text{MeV}}
\newcommand{\gev}{\text{GeV}}
\newcommand{\tev}{\text{TeV}}

\newcommand{\cm}{\text{cm}}
\newcommand{\m}{\text{m}}
\newcommand{\km}{\text{km}}
\newcommand{\mrad}{\text{mrad}}

\newcommand{\ps}{\text{ps}}

\newcommand{\eg}{{\em e.g.}}
\newcommand{\ie}{{\em i.e.}}

\renewcommand{\eqref}[1]{Eq.~(\ref{#1})}

\newcommand{\secref}[1]{Sec.~\ref{sec:#1}}
\newcommand{\secsref}[2]{Secs.~\ref{sec:#1} and \ref{sec:#2}}

\newcommand{\figref}[1]{Fig.~\ref{fig:#1}}

\newcommand{\lmin}{L_{\text{min}}}
\newcommand{\lmax}{L_{\text{max}}}
\newcommand{\mphi}{m_{\phi}}
\newcommand{\nsig}{N_{\text{sig}}}

\begin{document}

\preprint{UCI-TR-2017-12}

\title{\PRE{\vspace*{1.0in}}
Dark Higgs Bosons at FASER
\PRE{\vspace*{.4in}}}

\author{Jonathan L.~Feng}
\email{jlf@uci.edu}
\affiliation{Department of Physics and Astronomy, University of
California, Irvine, CA 92697-4575 USA
\PRE{\vspace*{.1in}}}

\author{Iftah Galon}
\email{iftachg@uci.edu}
\affiliation{Department of Physics and Astronomy, University of
California, Irvine, CA 92697-4575 USA
\PRE{\vspace*{.1in}}}
\affiliation{New High Energy Theory Center \\
Rutgers, The State University of New Jersey \\
Piscataway, New Jersey 08854-8019, USA
\PRE{\vspace*{.1in}}}

\author{Felix Kling}
\email{fkling@uci.edu}
\affiliation{Department of Physics and Astronomy, University of
California, Irvine, CA 92697-4575 USA
\PRE{\vspace*{.1in}}}

\author{Sebastian Trojanowski\PRE{\vspace*{.2in}}}
\email{strojano@uci.edu}
\affiliation{Department of Physics and Astronomy, University of
California, Irvine, CA 92697-4575 USA
\PRE{\vspace*{.1in}}}
\affiliation{National Centre for Nuclear Research,\\Ho{\. z}a 69, 00-681 Warsaw, Poland
\PRE{\vspace*{.4in}}}

%\date{\today}

\begin{abstract}
\PRE{\vspace*{.2in}} 
FASER, ForwArd Search ExpeRiment at the LHC, has been proposed as a small, very far forward detector to discover new, light, weakly-coupled particles.  Previous work showed that with a total volume of just $\sim 0.1 - 1~\m^3$, FASER can discover dark photons in a large swath of currently unconstrained parameter space, extending the discovery reach of the LHC program.  Here we explore FASER's discovery prospects for dark Higgs bosons.  These scalar particles are an interesting foil for dark photons, as they probe a different renormalizable portal interaction and are produced dominantly through $B$ and $K$ meson decays, rather than pion decays, leading to less collimated signals.  Nevertheless, we find that FASER is also a highly sensitive probe of dark Higgs bosons with significant discovery prospects that are comparable to, and complementary to, much larger proposed experiments.  
\end{abstract}

%\pacs{}

\maketitle

%%%%%%%%%%%%%%%%%%%%%%%%%%%%%
% Introduction
%%%%%%%%%%%%%%%%%%%%%%%%%%%%%
\section{Introduction}
\label{sec:introduction}

At present, no particles beyond the standard model (SM) have been found at the Large Hadron Collider (LHC).  So far, attention has typically focused on hypothetical heavy particles with SM gauge interactions, which give rise to high $p_T$ signatures, and the ATLAS and CMS experiments are optimized for such searches.  But light particles with milli-charged and weaker couplings are increasingly motivated (see, \eg, Ref.~\cite{Battaglieri:2017aum}), and are predominantly produced with low $p_T$.  
Such particles may have escaped detection at the LHC because they pass
undetected down the beam pipe, are long-lived and decay after leaving
existing detectors, are produced in the central region, but their signals are clouded by large SM background, or some combination of these possibilities.

In a previous paper~\cite{Feng:2017uoz}, we proposed that a new experiment, FASER (ForwArd Search ExpeRiment), be placed in the far forward region of either the ATLAS or CMS detector regions with the goal of discovering such new, light, weakly-coupled particles.  We considered two representative on-axis locations: a near location between the beampipes, after the neutral target absorber (TAN, or TAXN in the HL-LHC era), and roughly 135 m downstream; and a far location after the beamlines enter an arc, 400 m downstream.  In both locations, we found that a small cylindrical detector (4 cm in radius and 5 m deep in the near location, 20 cm in radius and 10 m deep in the far location) has significant discovery potential for new light particles.  As an example, we considered dark photons and found that for masses $m_{A'} \sim 10-500~\mev$ and micro- to milli-charged couplings ($\epsilon \sim 10^{-6} - 10^{-3}$), FASER could discover dark photons in a wide swath of currently unconstrained parameter space, with comparable sensitivity to other, much larger, proposed experiments.

In this study, we consider FASER's discovery potential for dark Higgs bosons.  As with dark photons that interact with the SM through a kinetic mixing term, dark Higgs bosons probe one of the few possible renormalizable interactions with a hidden sector, the Higgs portal quartic scalar interaction~\cite{Patt:2006fw}. In addition, dark Higgs bosons have numerous cosmological implications.  Like dark photons, they may mediate interactions with hidden dark matter (DM) that has the correct thermal relic density~\cite{Feng:2008ya} or resolves small scale structure discrepancies~\cite{Tulin:2017ara}.  Additionally, a dark Higgs boson may be the inflaton, providing a rare possibility to probe inflation in particle physics experiments~\cite{Shaposhnikov:2006xi,Bezrukov:2009yw,Bezrukov:2013fca,Bramante:2016yju}.  

From an experimental perspective, dark Higgs bosons are an interesting foil for dark photons.  Dark Higgs bosons mix with the SM Higgs boson and so inherit the property of coupling preferentially to heavy particles.  Dark Higgses are therefore dominantly produced in $B$ and $K$ meson decays, in contrast to dark photons, which are primarily produced in $\pi^0$ and light meson decays.  As a consequence, dark Higgs bosons are produced with greater $p_T$ and are less collimated, providing a challenging test case for FASER.  In addition, the trilinear scalar coupling $h \phi \phi$, where $h$ is the SM Higgs boson and $\phi$ is the dark Higgs bosons, can be probed both at FASER, through the double dark Higgs process $b \to s h^* \to s \phi \phi$, and through searches for the exotic SM Higgs decays $h \to \phi \phi$.  We will evaluate FASER's sensitivity to both $\phi$--$h$ mixing and the $h \phi \phi$ coupling, and compare them to other current and proposed experiments, such as NA62~\cite{NA62}, SHiP~\cite{Alekhin:2015byh}, MATHUSLA~\cite{Chou:2016lxi,Curtin:2017izq,Evans:2017lvd}, and CODEX-b~\cite{Gligorov:2017nwh}.  Throughout this study, we consider FASER in the high luminosity era and assume an integrated luminosity of $3~\iab$ at the 13 TeV LHC.

This study is organized as follows. In \secref{darkhiggs} we discuss dark Higgs bosons and their properties.  In \secsref{mixing}{trilinear} we determine FASER's sensitivity to $\phi$--$h$ mixing and the trilinear $h\phi \phi$ coupling, respectively.  We then note interesting implications for DM and inflation in \secref{cosmology} and present our conclusions in \secref{conclusions}.

%%%%%%%%%%%%%%%%%%%%%%%%%%%%%
% Dark Higgs
%%%%%%%%%%%%%%%%%%%%%%%%%%%%%
\section{Dark Higgs Properties}
\label{sec:darkhiggs}

If the SM is extended to include a hidden real scalar field $h'$, the most general scalar Lagrangian is
\begin{equation}
{\cal L} = \mu_H^2 |H|^2 - \frac{1}{4} \lambda_H |H|^4 
+\mu'^2 h'^2 - \mu'_3 h'^3 - \frac{1}{4} \lambda' h'^4 - \mu'_{12} h' |H|^2 - \epsilon h'^2 |H|^2 \ ,
\label{Lagrangian}
\end{equation}
where $H$ is the SM electroweak Higgs doublet, and the last term is the Higgs portal quartic scalar interaction.  To determine the physical particles and their properties, one must minimize the scalar potential and diagonalize the mass terms.  The resulting physical particles are a SM-like Higgs particle $h$ and a dark Higgs boson $\phi$.  The parameters are constrained by the SM-like Higgs boson's vacuum expectation value (vev) $v \simeq 246~\gev$ and mass $m_h \simeq 125~\gev$, but in general, five free parameters remain.  The number of free parameters can be reduced in specific models, for example, by invoking a discrete symmetry for $h'$ to set $\mu'_3 = \mu'_{12} = 0$, or by invoking such a discrete symmetry and further setting $\mu_H = 0$~\cite{Bezrukov:2009yw,Bezrukov:2013fca} or $\mu'=0$~\cite{Bramante:2016yju} by hand. 

For our purposes, it is most convenient to adopt a phenomenological parametrization, where the Lagrangian for the physical dark Higgs boson $\phi$ is
\begin{equation}
{\cal L} = - \mphi^2 \phi^2 - \sin \theta \, \frac{m_f}{v} \, \phi \bar{f} f  - \lambda  v h \phi \phi  + \ldots \ ,
\label{eq:Lphysical}
\end{equation}
where the omitted terms include additional cubic and quartic scalar interactions involving $\phi$ and $h$.  Current experimental constraints require $\sin \theta \approx \theta \ll 1$ and $\lambda \ll 1$.  We will refer to the three parameters, 
\begin{equation}
m_{\phi} \, , \ \theta , \  \lambda \ ,
\end{equation}
as the dark Higgs boson mass, mixing angle, and trilinear coupling, respectively.  They determine all of the phenomenological properties of interest here and will be taken as independent parameters throughout this study.

\subsection{Dark Higgs Decays}

The dark Higgs decay widths are suppressed by $\theta^2$ relative to those of a SM Higgs boson with identical mass. We will assume that there are no hidden sector decay modes.  For $\mphi < 2m_\pi$, then, the dark Higgs decays primarily to either $e^+ e^-$ or $\mu^+ \mu^-$ with decay width 
\be
\Gamma(\phi  \to \ell \ell) 
&= \frac{ m_\ell^2 \mphi}{8 \pi v^2} \left(1-\frac{4m_\ell^2}{\mphi^2}\right)^{3/2} \theta^2 \ ,
\ee
where $\ell = e, \mu$. In the mass range $2m_\pi < \mphi \alt 2.5~\gev$, the decay widths are complicated by decays to mesons and the effects of resonances, and there is no consensus regarding their values in the literature~\cite{Clarke:2013aya}. We adopt the numerical results of Ref.~\cite{Bezrukov:2013fca}, which incorporate the results of Ref.~\cite{Donoghue:1990xh} for the mass range $2 m_{\pi} < \mphi \alt 1~\gev$, use the spectator model~\cite{Gunion:1989we,McKeen:2008gd} for the high-mass range $\mphi \agt 2.5~\gev$, and interpolate between these two for the intermediate mass range $1~\gev \alt \mphi \alt 2.5~\gev$. 

\begin{figure}[tb]
\centering
\includegraphics[width=0.47\textwidth]{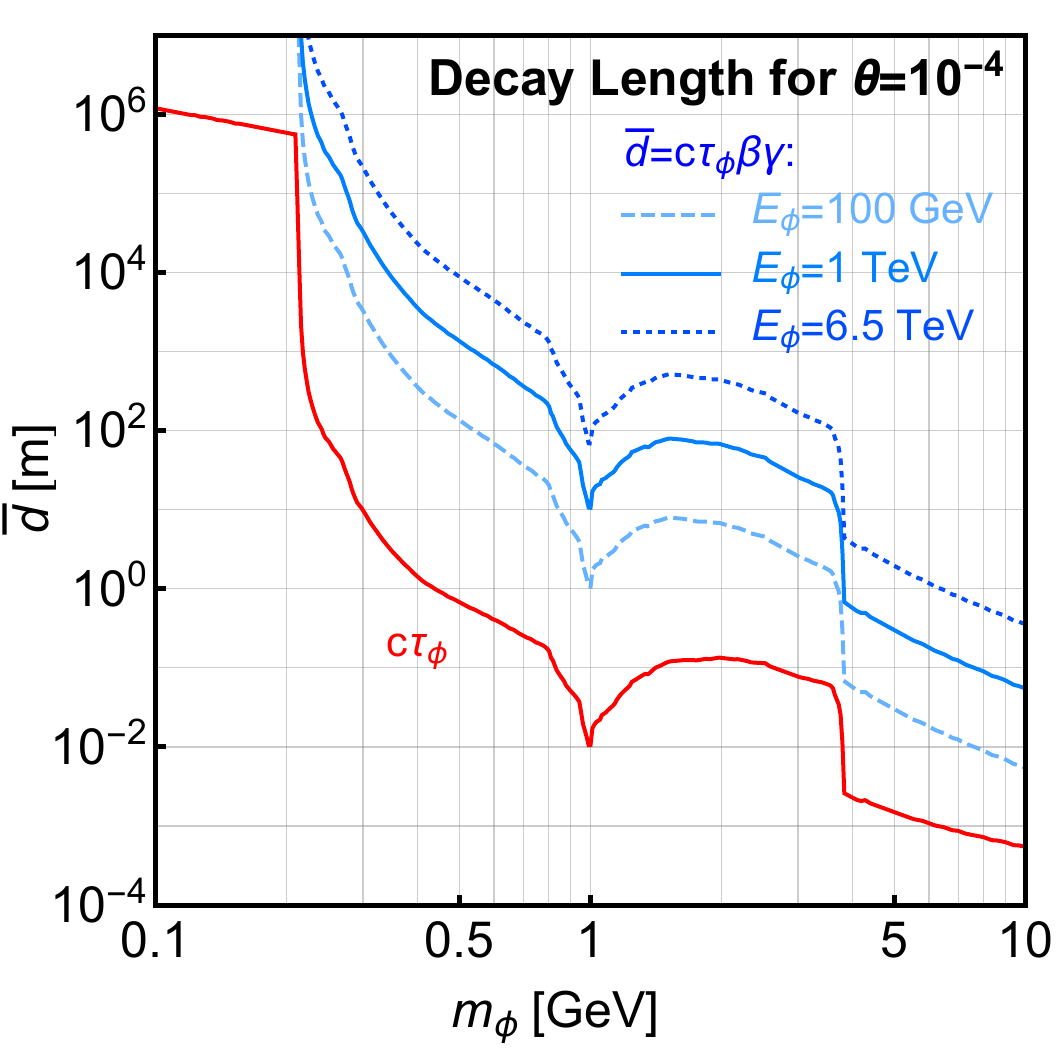} \quad 
\includegraphics[width=0.47\textwidth]{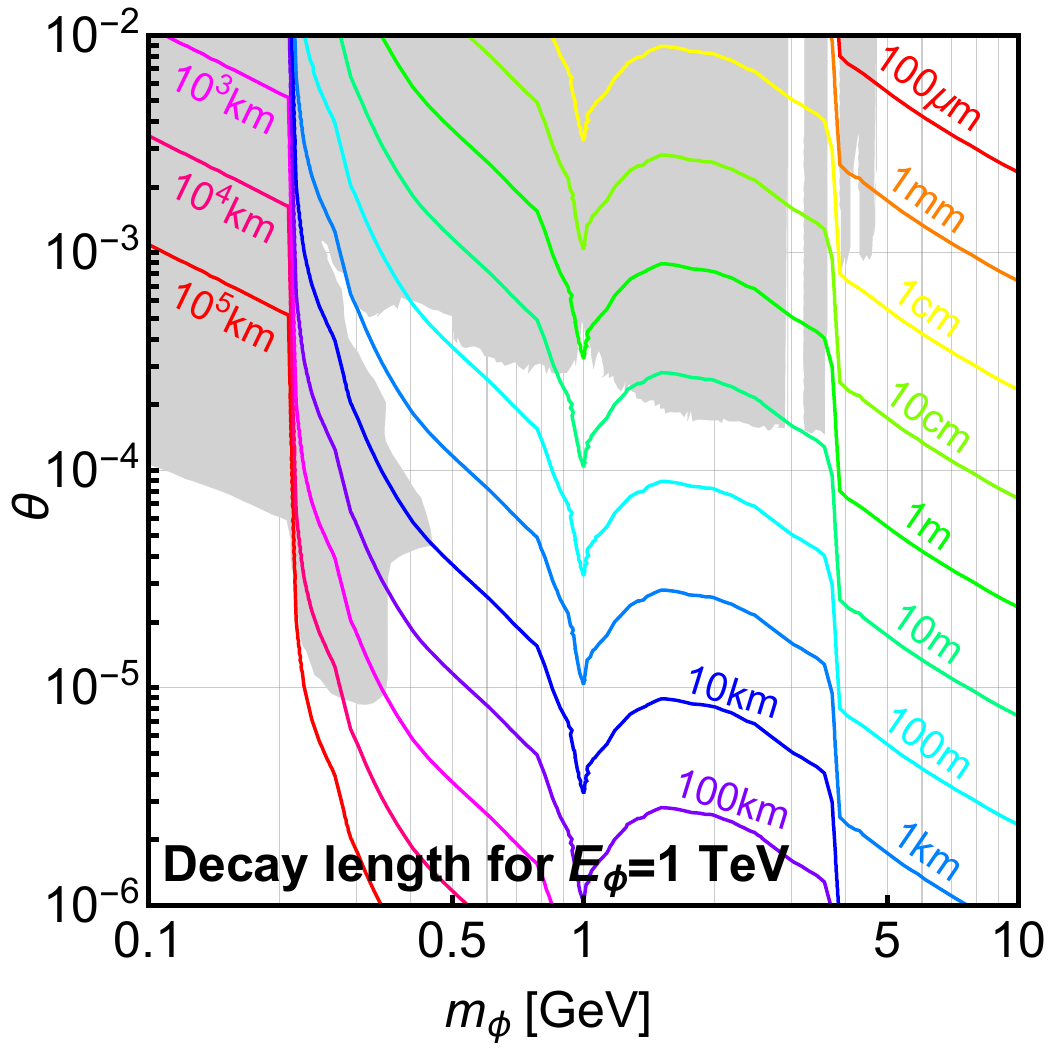} 
\caption{Left: Dark Higgs decay length $\bar{d} = c \tau_{\phi} \beta \gamma$ as a function of $\mphi$ for various energies $E_\phi$ and $\theta = 10^{-4}$.  The decay length scales as $\bar{d} \propto \theta^{-2}$. Adapted from Ref.~\cite{Bezrukov:2013fca}.  Right: Dark Higgs decay length $\bar{d}$ in the $(\mphi, \theta)$ plane for $E_{\phi} = 1~\tev$. The decay length scales as $\bar{d} \propto E_{\phi}$ for large $E_{\phi}$. The gray shaded regions are experimentally excluded. }
\label{fig:decays}
\end{figure}

The resulting decay lengths are shown in \figref{decays}. Because the decays are both Yukawa- and $\theta$-suppressed, for currently viable values of $\theta$ and energies $E_{\phi} \sim 1~\tev$, dark Higgs decay lengths can be very long.   Below the muon threshold, \ie, for $\mphi < 2 m_{\mu}$, the tiny electron Yukawa coupling leads to an extremely long dark Higgs lifetime, resulting in a negligible event rate in FASER, as most dark Higgs bosons typically overshoot the detector. On the other hand, for $2 m_{\mu} < \mphi \alt 2 m_{\tau}$ and energies $E_\phi \sim 100~\gev - 1~\tev$, decay lengths $\bar{d} = c \tau_\phi \beta \gamma \sim 10~\m - 1~\km$ are possible, and a significant number of dark Higgs bosons can pass through many LHC infrastructure components and decay within the FASER volume. 
 
\begin{figure}[tb]
\centering
\includegraphics[width=0.47\textwidth]{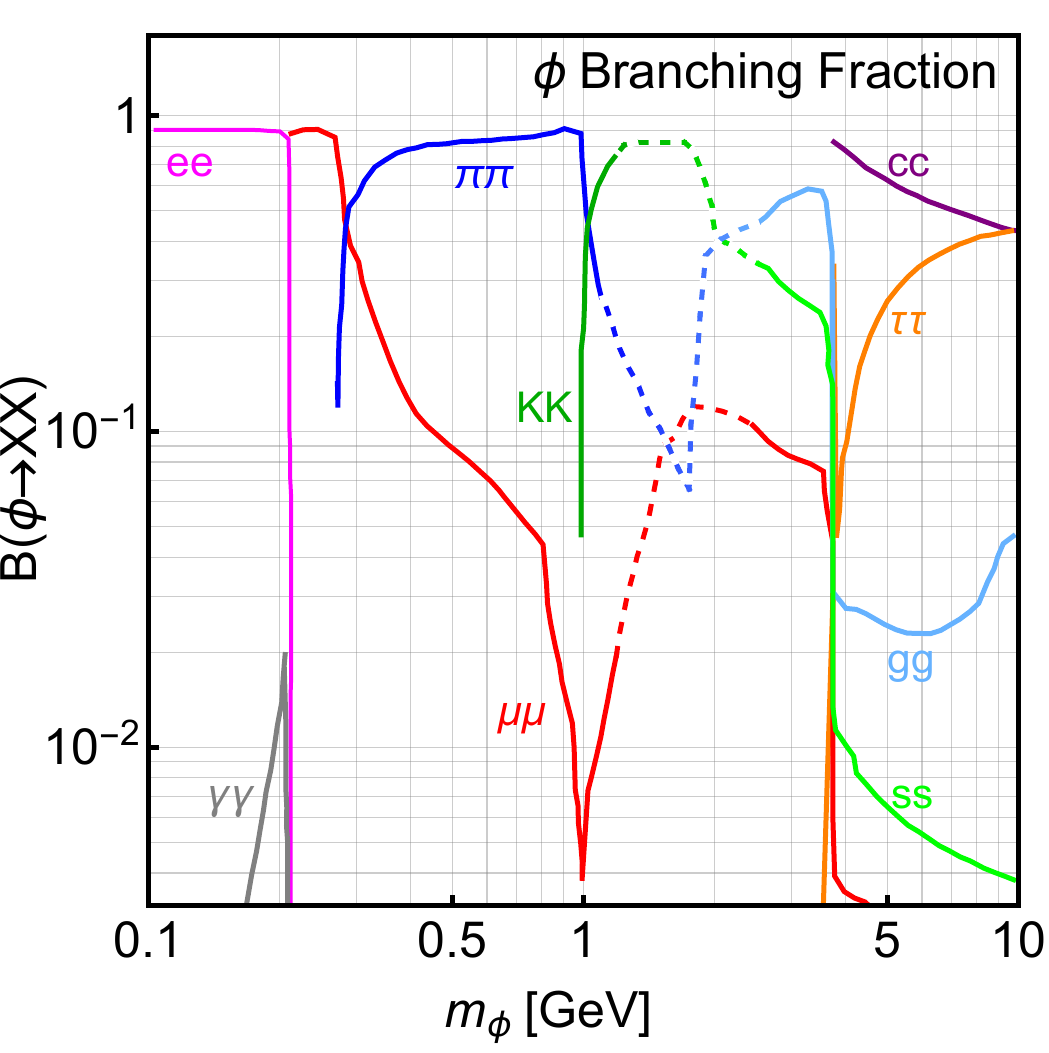} 
\caption{Dark Higgs branching fractions as a function of $\mphi$.  Adapted from Ref.~\cite{Bezrukov:2013fca}. }
\label{fig:bf}
\end{figure}

The dark Higgs branching fractions are shown in \figref{bf}.  As shown there, above the muon threshold, the $\mu^+ \mu^-$ decay mode dominates in the narrow region $2m_{\mu} < \mphi < 2m_{\pi}$, but for larger masses, the dominant decay modes are to pions, kaons, and other hadrons. This differs markedly from the dark photon case, where leptonic decays are significant for most of the mass range.

\subsection{Dark Higgs Production}

In this section, we discuss the dominant production mechanisms for dark Higgs bosons: $B$, $K$, and light meson decays.

Dark Higgs bosons can also be produced through other processes.  For example, they may be radiated off a $b$ quark line and be produced in processes $b \bar{b} \to \phi g$ or $b g \to b \phi$, or through the vector boson fusion processes $q q \to q q \phi$.\footnote{The contributions from processes such as gluon-gluon fusion cannot be reliably estimated, since the parton distribution functions suffer from large uncertainties when the light dark Higgs is produced in the forward direction~\cite{Feng:2017uoz}.}  In principle, such processes could extend the reach of FASER to masses $\mphi > m_B$.  We have checked, however, that, for currently viable values of dark Higgs parameters, such processes do not contribute significantly to dark Higgs rates in FASER, and so we focus instead on the meson decay processes in this study.

\subsubsection{$B$ Decays}

Single dark Higgs bosons may be produced in meson decays through $\phi$--$h$ mixing.  The rates are proportional to $\theta^2$ and, since the dark Higgs inherits the couplings of the SM Higgs, the branching ratios are largest for processes involving heavy flavors, in particular, $B$ mesons.

The inclusive decay of $B$ mesons into dark Higgs bosons is dominated by the parton-level process $b \to s \phi$ containing a $t$--$W$ loop, with the $\phi$ radiated from the top quark. Uncertainties from strong interaction effects are minimized in the ratio~\cite{Grinstein:1988yu,Chivukula:1988gp}
\be
\frac{\Gamma(B \to X_s \phi)}{\Gamma(B \to X_c e \nu) }= \frac{\Gamma(b \to s \phi )}{\Gamma(b \to c e \nu)} = \frac{27}{64 \pi^2 v^2} \frac{m_t^4}{m_b^2 } \left(1-\frac{\mphi^2}{m_b^2}\right)^2 \frac{1}{f_{c/b}} \left| \frac{V_{ts}^* V_{tb}}{V_{cb}}\right|^2\; \theta^2 \ ,
\ee
where $X_{s,c}$ denote any strange and charm hadronic state, and $f_{c/b} \simeq 0.51$. Given $B(B \to X_c e \nu)=0.104$ for both $B^0$ and $B^\pm$, and since the total width is $\theta$-independent for $\theta \ll 1$, we find
\be
B(B \to X_s \phi) \simeq 5.7 \left(1-\frac{\mphi^2}{m_b^2}\right)^2 \theta^2 \ .
\label{eq:B_BtoX_sphi}
\ee
Here and below we take the quark masses to be $m_t \simeq 173~\gev$, $m_b \simeq 4.75~\gev$, $m_s \simeq 95~\mev$, and we use the most recent values of meson decay widths and CKM parameters~\cite{Patrignani:2016xqp}.  In the following, we apply \eqref{eq:B_BtoX_sphi} to obtain the $\phi$ production rate in $B$ meson decays, which we regard as a two-body decay. In doing so, we neglect small kinematic effects which may arise when $X_s$ is a multi-body state.

\subsubsection{$K$ and Light Meson Decays}

The amplitude for $K$ decay into a dark Higgs boson is~\cite{Leutwyler:1989xj,Gunion:1989we, Bezrukov:2009yw}
\be
\mathcal{M} = \theta \frac{m_K^2}{v} \left( \gamma_1 \, \frac{7}{18}\frac{m_K^2-\mphi^2+m_\pi^2}{m_K^2}-\gamma_2 \, \frac{7}{9} + \frac{1}{2} \frac{3 }{16 \pi^2 v^2} \sum_{i=u,c,t}V^*_{id}m_i^2V_{is}\right) \ ,
\ee
where $\gamma_1=3.1\times 10^{-7}$ and $\gamma_2 \approx 0$. The third term is from $(u, c, t)$--$W$ loop diagrams, and it is again dominated by the top contribution. The branching fractions for the physical kaon states are
\be
B(K^\pm \to \pi^\pm \phi)
&= \frac{1}{\Gamma_{K^\pm}} \frac{2 p_\phi^0}{m_K} \frac{|\mathcal{M}|^2}{16 \pi m_K} 
=  2.0 \times 10^{-3} \;\frac{2 p_\phi^0}{m_K}\; \theta^2 \\
B(K_L \to \pi^0 \phi)
&= \frac{1}{\Gamma_{K_L}} \frac{2 p_\phi^0}{m_K} \frac{\text{Re}(\mathcal{M})^2}{16 \pi m_K} 
=  7.0 \times 10^{-3}\; \frac{2 p_\phi^0}{m_K}\; \theta^2 \\
B(K_S \to \pi^0 \phi)
&= \frac{1}{\Gamma_{K_S}} \frac{2 p_\phi^0}{m_K} \frac{\text{Im}(\mathcal{M})^2}{16 \pi m_K} 
=  2.2 \times 10^{-6}\; \frac{2 p_\phi^0}{m_K}\; \theta^2 \ ,
\ee
where $p_\phi^0 = \lambda^{1/2} (m_K^2, m_{\pi}^2, m_{\phi}^2) / (2 m_K)$, with $\lambda(a,b,c) = a^2+b^2+c^2- 2(a b + a c + b c)$, is the dark Higgs boson's three-momentum in the parent meson's rest frame.  The ratio $2p_{\phi}^0/m_K$ is unity for $m_{\pi}, \mphi \ll m_K$.  $B(K_S \to \pi^0 \phi)$ is suppressed relative to the others primarily by the relatively large $K_S$ total decay width, but also by the small $CP$-violating phase in the CKM matrix.  
%\Gamma_{K^\pm}=5.317 \cdot 10^{-17}~\gev 
%\Gamma_{K_L}=1.287 \cdot 10^{-17}~\gev
%\Gamma_{K_S}=7.351 \cdot 10^{-15}~\gev

Dark Higgs bosons may also be produced in the decays of light mesons, for example, through processes with branching fractions   $B(\eta' \to \eta \phi) \simeq 7.2 \times 10^{-5} (2 p_\phi^0 / m_{\eta'}) \theta^2$~\cite{Gunion:1989we}, $B(\eta \to \pi^0 \phi) \sim  10^{-6} (2 p_\phi^0 / m_{\eta}) \theta^2$~\cite{Leutwyler:1989xj,Kozlov:1995yd}, and $B(\pi^\pm \to e \nu \phi) \simeq 1.9 \times 10^{-9} f(\mphi^2 / m_\pi^2) \theta^2$, where $f(x)=(1-8x+x^2)(1-x^2)-12x^2\ln x$~\cite{Dawson:1989kr}.

As expected, the branching fractions have the hierarchy $B(B \to \phi) \gg B(K \to \phi) \gg B(\eta, \pi \to \phi)$. The numbers of kaons and light mesons produced at the LHC are very roughly comparable, and so, given the hierarchy in branching fractions, kaon decay is always a more effective production mechanism for dark Higgs bosons than light meson decay.  The number of $B$ mesons produced at the LHC is, of course, suppressed relative to kaons, but the larger branching fraction compensates for this, and also $B$ decays probe much higher $\mphi$.  Given these considerations, we will show results for $B$ and $K$ decays below, and neglect those for light meson decays.

%%%%%%%%%%%%%%%%%%%%%%%%%%%%%
% Probes of Dark Higgs--SM Higgs Mixing
%%%%%%%%%%%%%%%%%%%%%%%%%%%%%
\section{Probes of Dark Higgs-SM Higgs Mixing}
\label{sec:mixing}

Given the production and decay properties of dark Higgs bosons described in \secref{darkhiggs}, we now determine the sensitivity of FASER to dark Higgs bosons produced through $\phi$--$h$ mixing.  In \secref{mesonproduction} we describe the parent meson and dark Higgs boson kinematic distributions at the LHC.  In \secref{insidedetector} we determine the number of dark Higgs bosons that decay in FASER, for various possible realizations of FASER, and in \secref{reach} we estimate the discovery potential for dark Higgs bosons in the $(\mphi, \theta)$ parameter space.

\subsection{Meson and Dark Higgs Boson Distributions}
\label{sec:mesonproduction}

To determine the dark Higgs boson event yield in FASER, we first simulate $B$ and $K$ production in the very far forward region at the LHC.  For kaons, we follow the procedure described in Ref.~\cite{Feng:2017uoz} and use the Monte-Carlo event generator EPOS-LHC~\cite{Pierog:2013ria}, as implemented in the CRMC simulation package~\cite{CRMC}. 

To simulate $B$ mesons, we use the simulation tool FONLL~\cite{Cacciari:1998it, Cacciari:2012ny, Cacciari:2015fta}, which calculates the differential cross section $d\sigma(pp\to B+X)/(dydp_T^2)$. This is obtained from a convolution of a perturbative partonic production cross section with a non-perturbative fragmentation function, which follows a Kartvelishvili et al.\ distribution with fragmentation parameter $\alpha=24.2$~\cite{Kartvelishvili:1977pi,Cacciari:2005uk}.  The dominant contribution to $B$ production comes from the parton-level process $gg \to b\bar{b}$.  The typical momentum transfer for this process in the far forward direction, where $\theta_b \ll 1$, is $q^2 \approx 2 m_b^2 + \frac{1}{2} p_{T,b}^2 \approx 50~\gev^2$. The partonic center-of-mass energy $\hat{s}=x_1 x_2 s$ is bounded from below by $\hat{s}>4m_b^2$. For the 13 TeV LHC, then, $b$-quark pair production receives contributions from momentum fractions $x_{1,2}$ as low as $4 m_b^2/ s \simeq  5 \times 10^{-7}$.  For momentum transfers $q^2 \sim 50~\gev^2$, the parton distribution functions (pdfs) are well behaved even at this low $x$, but suffer from uncertainties as large as a factor of 2.  We use the CTEQ 6.6 pdfs with $m_b=4.75~\gev$.  

\begin{figure}[tb]
\centering
\includegraphics[width=0.32\textwidth]{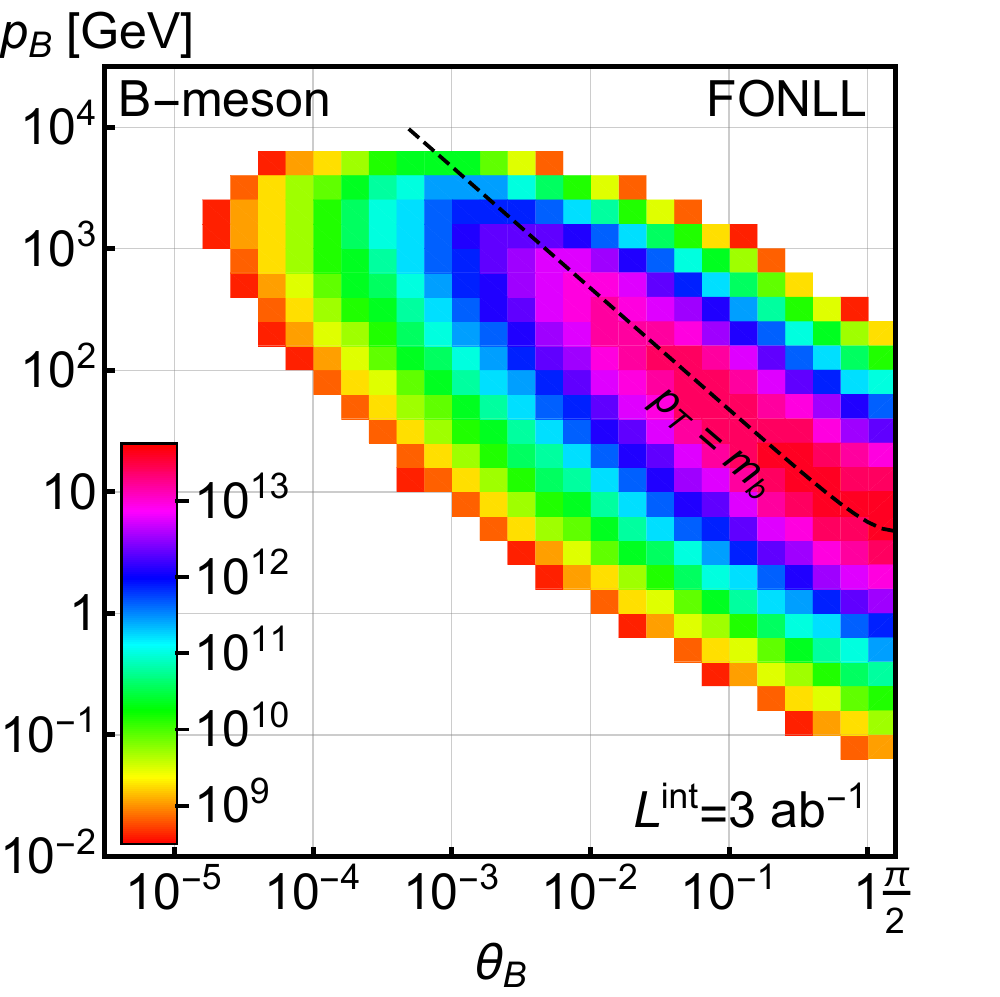}
\includegraphics[width=0.32\textwidth]{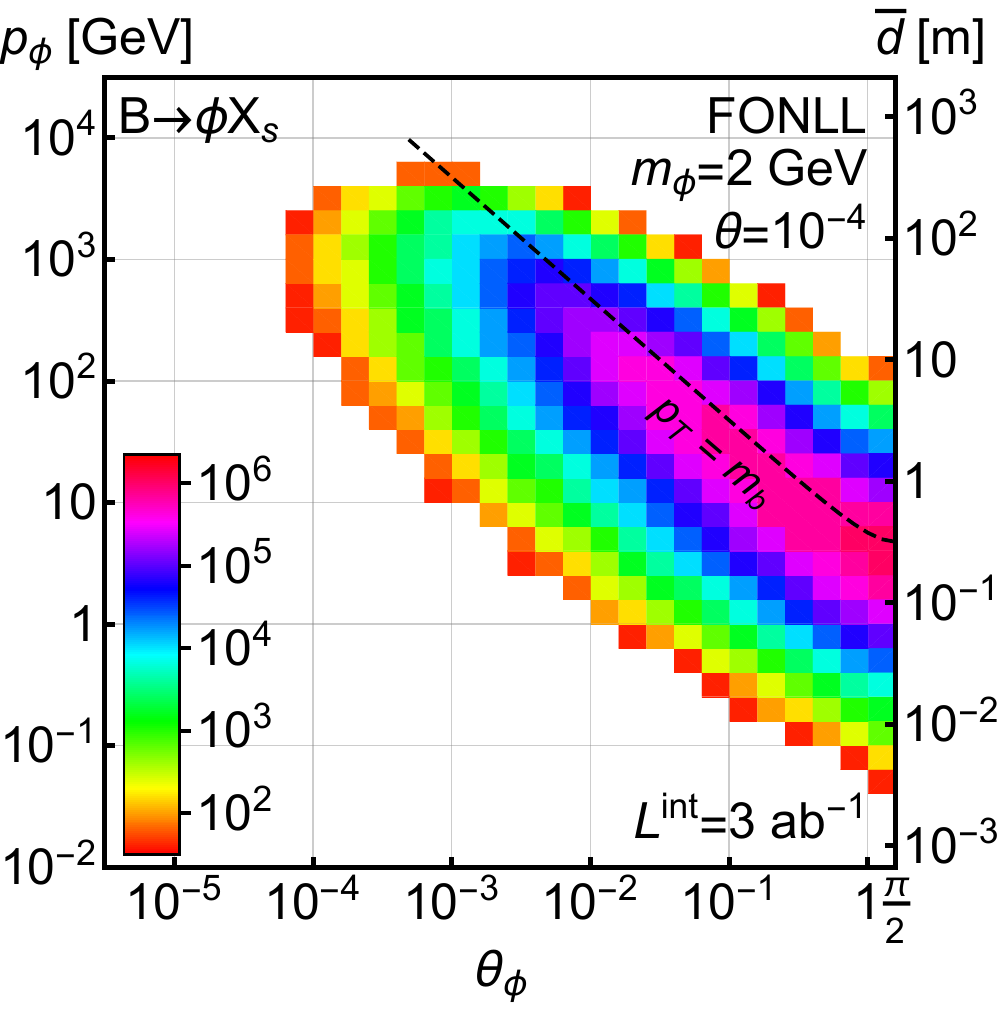}
\includegraphics[width=0.32\textwidth]{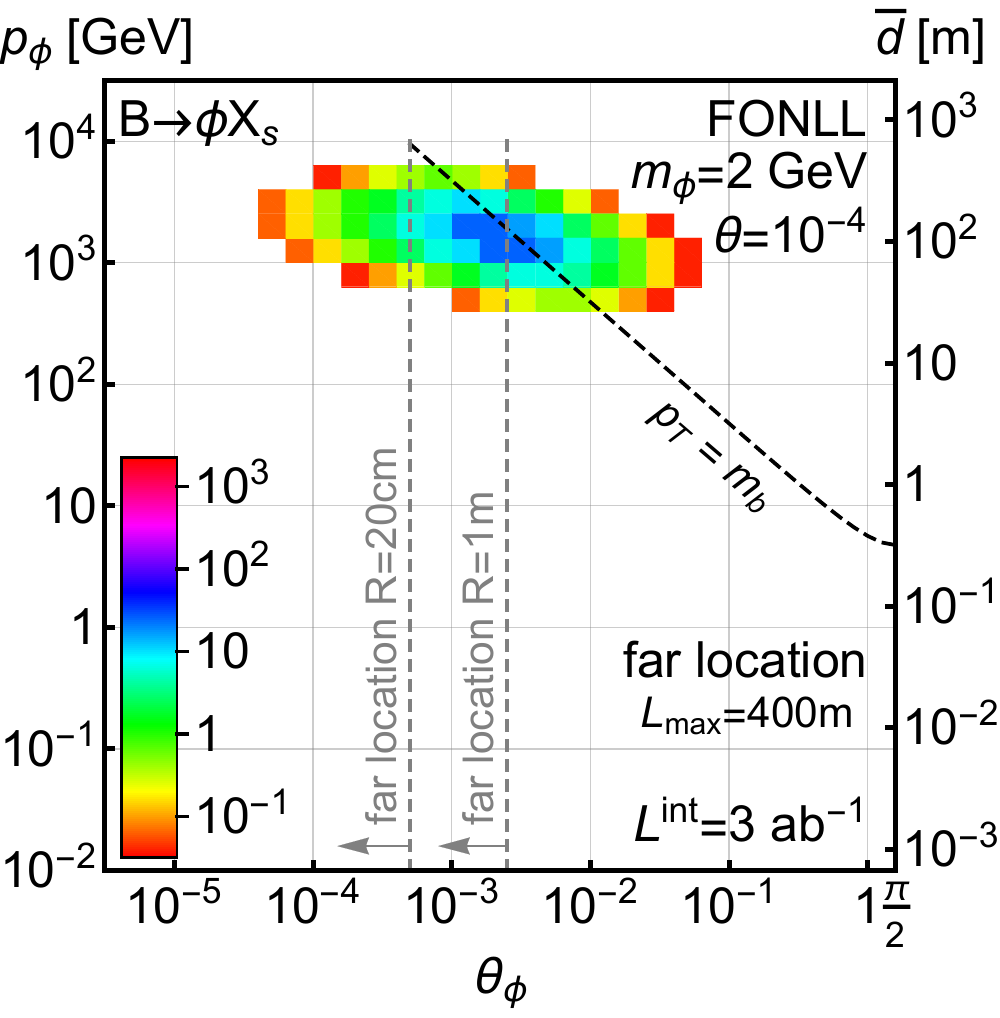}\\
\includegraphics[width=0.32\textwidth]{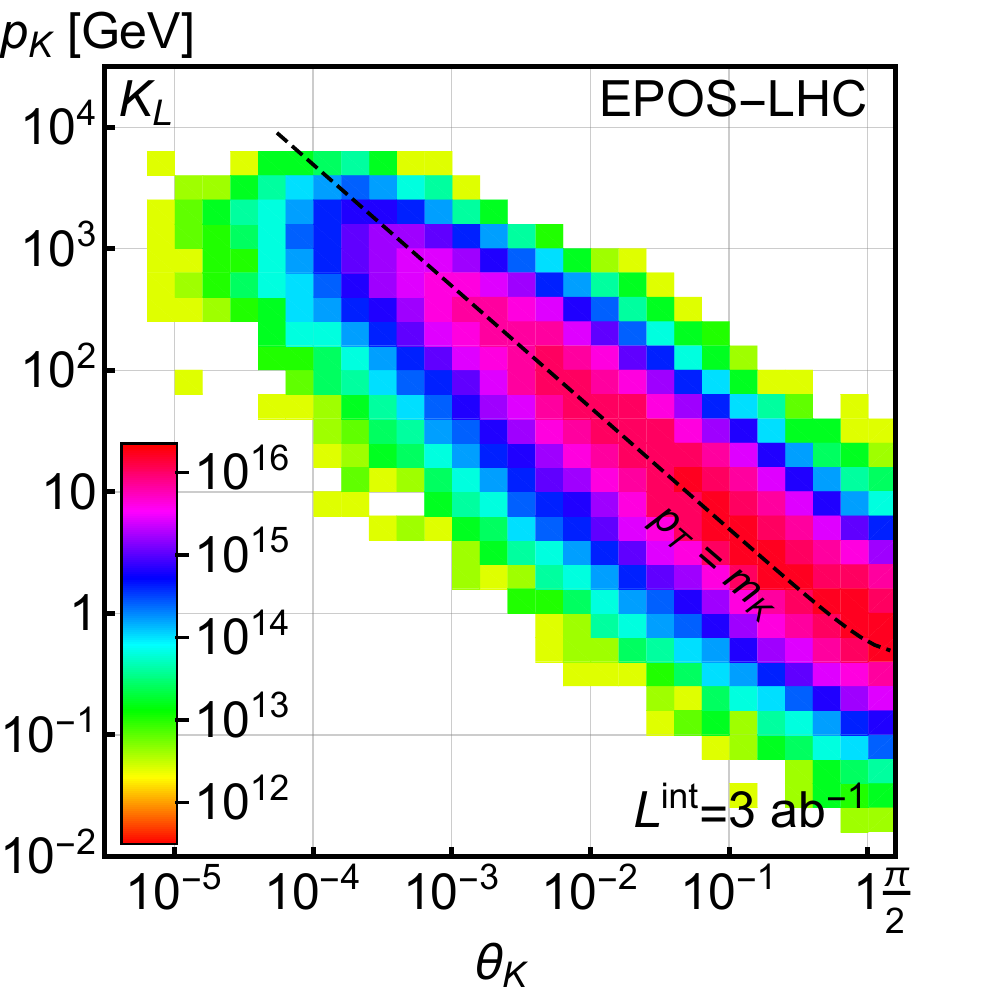}
\includegraphics[width=0.32\textwidth]{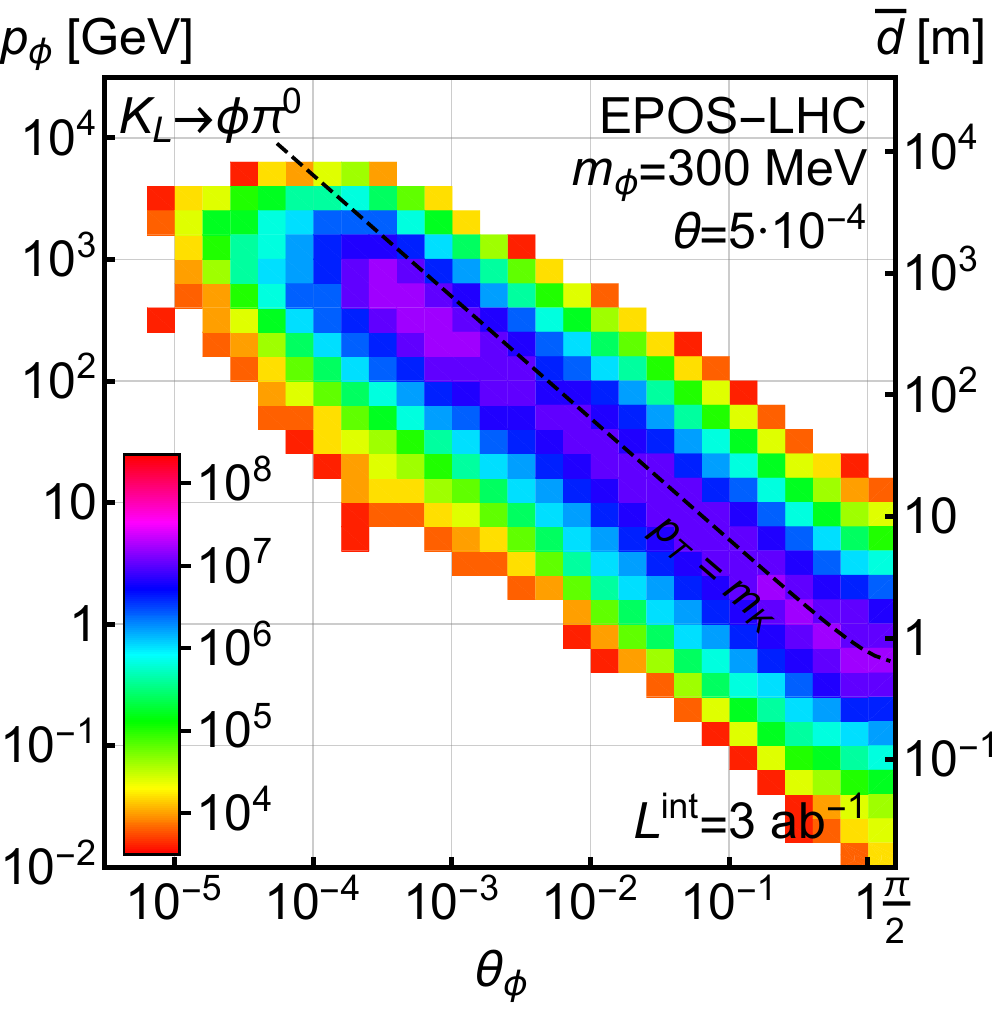}
\includegraphics[width=0.32\textwidth]{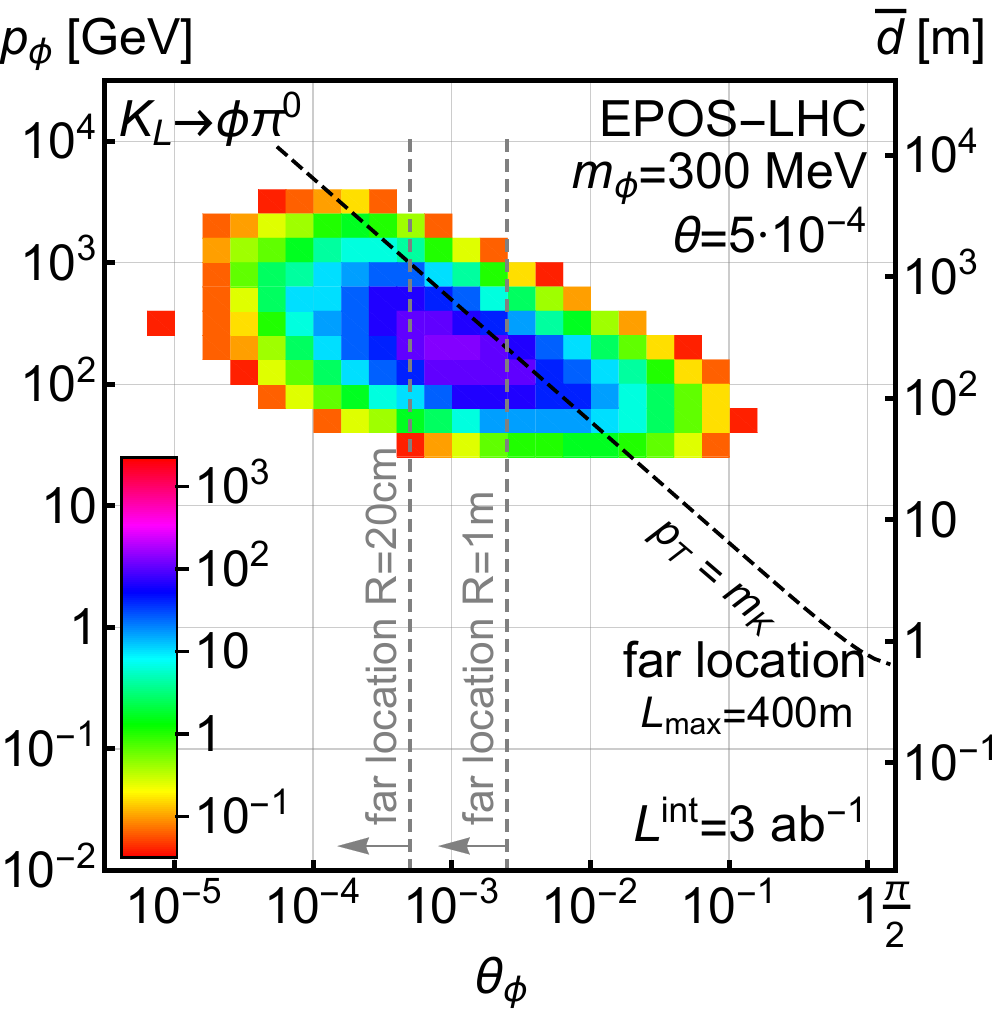}
\caption{Distribution of particles produced at the 13 TeV LHC with an integrated luminosity of $3~\iab$ in the $(\theta, p)$ plane, where $\theta$ and $p$ are the particle's angle with respect to the beam axis and momentum, respectively.  The panels show the number of particles produced in one hemisphere ($0 < \cos \theta \le 1$).  The bin thickness is $1/5$ of a decade along each axis.  The top row shows the distributions of $B$ mesons (left), dark Higgs bosons produced in $B$ decays (center), and dark Higgs bosons produced in $B$ decays that themselves decay after traveling a distance in the range $(\lmin, \lmax) = (390~\m, 400~\m)$ (right) for model parameters $(m_\phi, \theta) = (2~\gev, 10^{-4})$. The bottom row shows the analogous distributions for $K_L$ and $(m_\phi, \theta) = (300~\mev, 5 \times 10^{-4})$.  The black dashed lines corresponds to $p_T = p \sin \theta =  m_B$ in the top row and $m_K$ in the bottom row, and the gray dashed vertical lines in the right panels show the angular coverage of two representative configurations of FASER in the far location.
}
\label{fig:PvsT}
\end{figure}

In the top left and bottom left panels of \figref{PvsT}, we show the kinematic distributions of $B$ and $K_L$ mesons in the $(\theta_{B},p_{B})$ and $(\theta_{K},p_{K})$ planes, respectively, where $\theta_{B,K}$ and $p_{B,K}$ are the meson's angle with respect to the beam axis and momentum, respectively.  With an integrated luminosity of $3~\iab$, the 13 TeV LHC produces $1.42 \times 10^{15}$ $B$ mesons~\cite{Aaij:2016avz} and $9.1 \times 10^{17}$ kaons in each hemisphere, and these are clustered around $p_T \sim m_{B}$ and $m_K$, respectively. 

To derive the dark Higgs distributions from the meson distributions, we decay each $B$ and $K$ meson in the Monte Carlo sample.  $B$ mesons decay essentially at the IP; kaons travel macroscopic distances before decaying, and we scan over the kaon decay positions with the proper weighting.  We further consider only kaons that decay before colliding with the beampipe at a transverse distance of 3 cm, and restrict our attention to $K_{L,S}$ mesons that decay before reaching the TAN/TAXN at $L \approx 140~\m$ and to $K^{\pm}$ mesons that decay before being deflected by the Q1 magnet at $L=20~\m$. 

The dark Higgs boson distributions are shown in \figref{PvsT} for model parameters $(\mphi,\theta)=(2~\gev, 10^{-4})$ for $B$ decays (top center) and $(\mphi,\theta)=(300~\mev, 5\times 10^{-4})$ for kaons (bottom center). The dominant contribution from kaons is from $K_L$ decays,\footnote{The $K_L$ distribution is clustered around $p_T \sim m_K$, and along this line, there are two dominant populations with energies around 1-10 GeV and 100-1000 GeV. Those with intermediate energies are removed by the requirement that the kaon decay within the beampipe.} with a subleading contribution from $K^\pm$ decays.  The $K_S$ contribution is suppressed by a factor of 20 relative to the $K_L$ contribution; the short $K_S$ lifetime reduces the branching ratio to dark Higgs bosons, but also guarantees that all $K_S$ decay before hitting the TAXN. Despite the suppression of branching fractions by $\theta^2$, we see that, even for $\theta = 10^{-4}$, for an integrated luminosity of $3~\iab$ at the 13 TeV LHC, $B$ and $K$ decays each produce $\sim 10^6 - 10^7$ dark Higgs bosons that are boosted toward FASER.  

\subsection{Dark Higgs Decay Inside Detector}
\label{sec:insidedetector}

To determine the number of dark Higgs bosons that decay in FASER, we must specify the size, shape, and location of FASER.  As in Ref.~\cite{Feng:2017uoz}, we consider cylindrical detectors centered on the beam collision axis with radius $R$ and depth $\Delta = \lmax - \lmin$, where $\lmax$ ($\lmin$) is the distance from the IP to the far (near) edge of the detector along the beam axis. We consider two representative detector locations:
\be
\textbf{far location: }   &\lmax=400~\m,\, \Delta=10~\m,\, R=20~\cm, \, 1~\m  \\
\textbf{near location: } &\lmax=150~\m,\, \Delta=5~\m,\, R=4~\cm \ .
\label{detector-location}
\ee
The far detector is placed along the beam axis after the beam tunnel starts to curve. The near location is located between the beams and between the TAN/TAXN neutral particle absorber and the D2 dipole magnet.  The rationale for these locations is given in Ref.~\cite{Feng:2017uoz}.  Other locations between the near and far location (or even in an existing service tunnel slightly beyond the far location~\cite{Feng:2017uoz}) may be more feasible, but we will present results for these two as they represent two natural extremes.  Note that, in addition to the far detector with $R = 20~\cm$ studied in Ref.~\cite{Feng:2017uoz}, we also consider a larger detector with $R=1~\m$, for reasons that will become clear below.

In the case of dark Higgs bosons produced in $B$ decays, which typically occurs very close to the IP, the probability of a dark Higgs boson to decay inside the detector volume is
\be
\label{eq:P_decay_in_volume}
\mathcal{P}_{\phi}^{\text{det}} (p_{\phi},\theta_{\phi})
= ( e^{-\lmin/\bar{d}} - e^{-\lmax/\bar{d}} ) \
\Theta(R-\tan\theta_{\phi}\lmax) \ ,
\ee
where the first term is the probability that the dark Higgs boson decays within the $(\lmin,\lmax)$ interval, and the second term enforces the angular acceptance of the detector. For kaon decays, since kaons travel macroscopic distances before decaying, \eqref{eq:P_decay_in_volume} is modified to take into account both the horizontal and vertical displacement of the position at which the dark Higgs boson is produced. In the right panels of \figref{PvsT}, we show the number of dark Higgs bosons decaying in the far location range $(\lmin , \lmax) = (390~\m, 400~\m)$. We see that only very energetic particles with $E_\phi \agt 100~\gev$ have a sufficient decay length $\bar{d}$ to reach the detector. 

\begin{figure}[tb]
\centering
\includegraphics[width=0.47\textwidth]{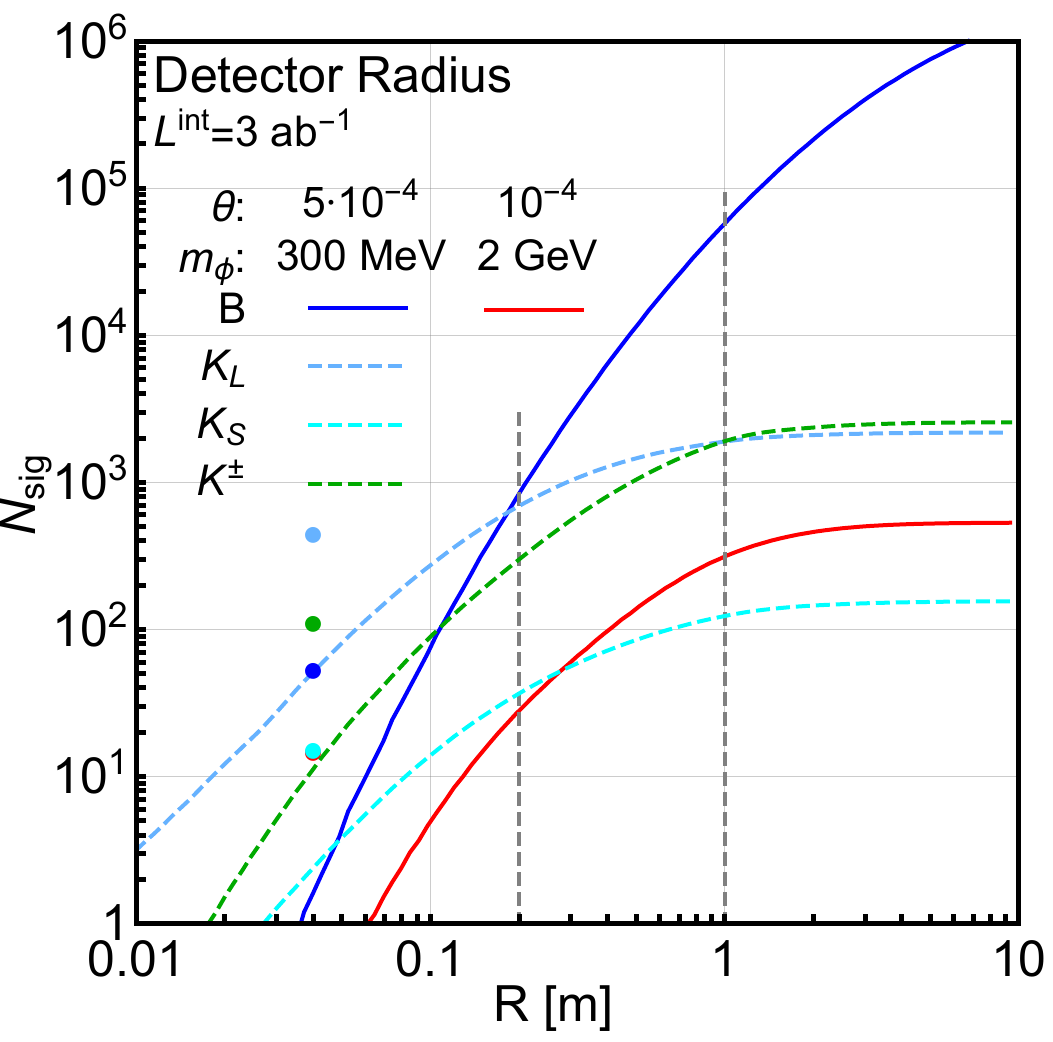}
\caption{Number of dark Higgs bosons produced at the 13 TeV LHC with $3~\iab$ that decay in FASER at the far location as a function of detector radius $R$.  Results are given for dark Higgs bosons produced in $B$, $K_L$, $K_S$, and $K^{\pm}$ decay for model parameters $(\mphi,\theta)= (300~\mev, 5\times 10^{-4})$, and for dark Higgs bosons produced in $B$ decay for model parameters $(2~\gev, 10^{-4})$.  The vertical lines at $R = 20~\cm$ and 1 m are the radii for two representative far location detectors.  Results for the near location with fixed $R = 4~\cm$ are also shown. 
}
\label{fig:radius}
\end{figure}

In \figref{radius} we explore the dependence of the signal rates at the far location on the detector radius $R$. We consider the two model parameter points selected previously: $(\mphi,\theta)=(2~\gev, 10^{-4})$ and $(300~\mev, 5\times 10^{-4})$, which correspond to $c \tau_\phi \approx 0.1~\m$ and $c \tau_\phi \approx 0.4~\m$, respectively. In the first scenario, $\mphi > m_K$, and so the dark Higgs production is predominantly through $B$ meson decays.  To reach FASER, the dark Higgs boson must have a large boost factor corresponding to energies above  $\sim 1~\tev$, as seen in the top right panel of \figref{PvsT}. Dark Higgs bosons with such high energies are already very collimated, and extending the detector radius from $R=20~\cm$ to $R=1~\m$ does not dramatically improve the signal acceptance, as seen in \figref{radius}. 

In the second case, the dark Higgs bosons is both lighter and longer lived, so the spectrum of dark Higgs bosons that can decay in FASER extends to lower energies. In contrast to the former scenario, for this benchmark, $\mphi < m_K$, so dark Higgs bosons are produced in both $K$ and $B$ decays. For the $B$ decays, lower energies imply that the signal is less collimated, and indeed, as can be seen in \figref{radius}, extending the detector radius from $R=20~\cm$ to $R=1~\m$ improves the signal event yield by two orders of magnitude. On the other hand, the kaon distributions are highly collimated as they are centered along $p_T \sim m_K$, and the effect on the signal from kaon decays is therefore negligible.

\subsection{Mixing Reach}
\label{sec:reach}

\begin{figure}[t]
\centering
\includegraphics[width=0.47\textwidth]{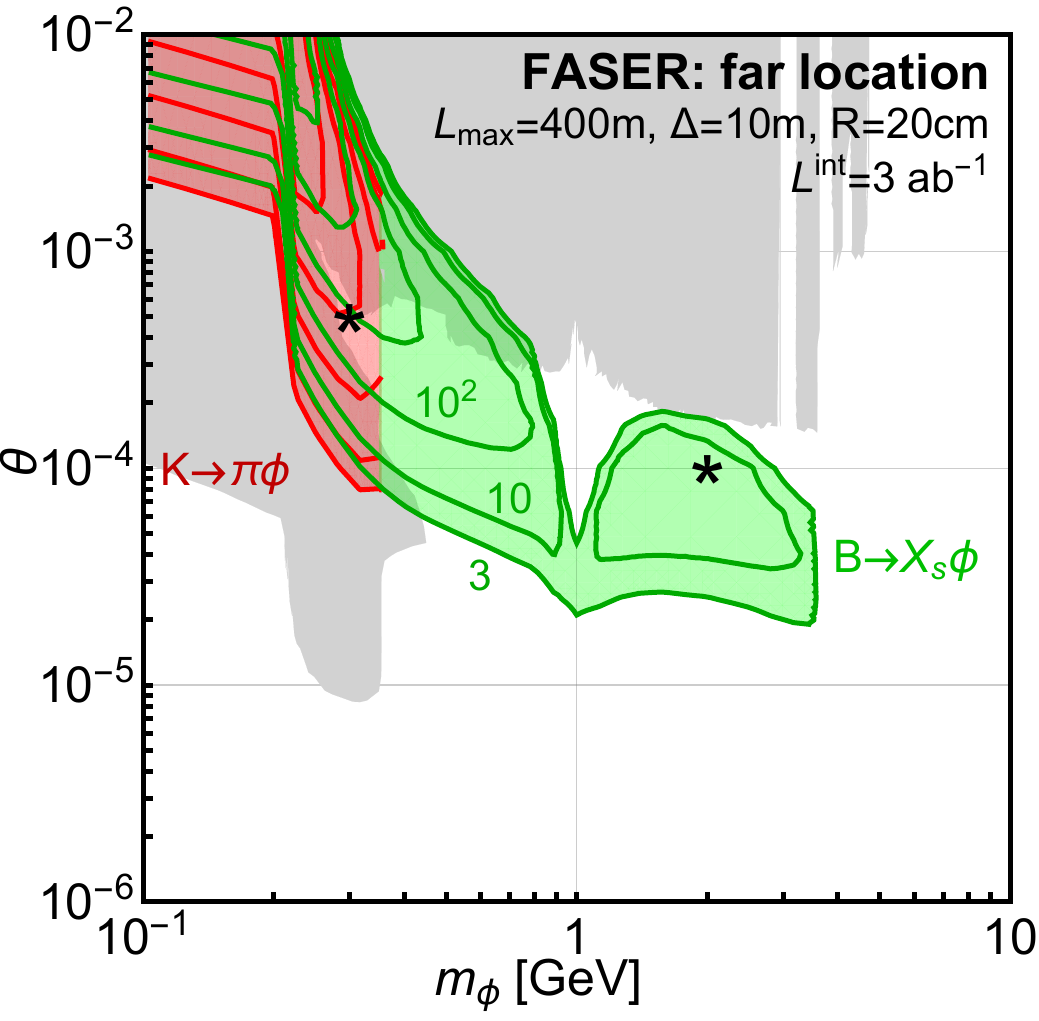} \quad 
\includegraphics[width=0.47\textwidth]{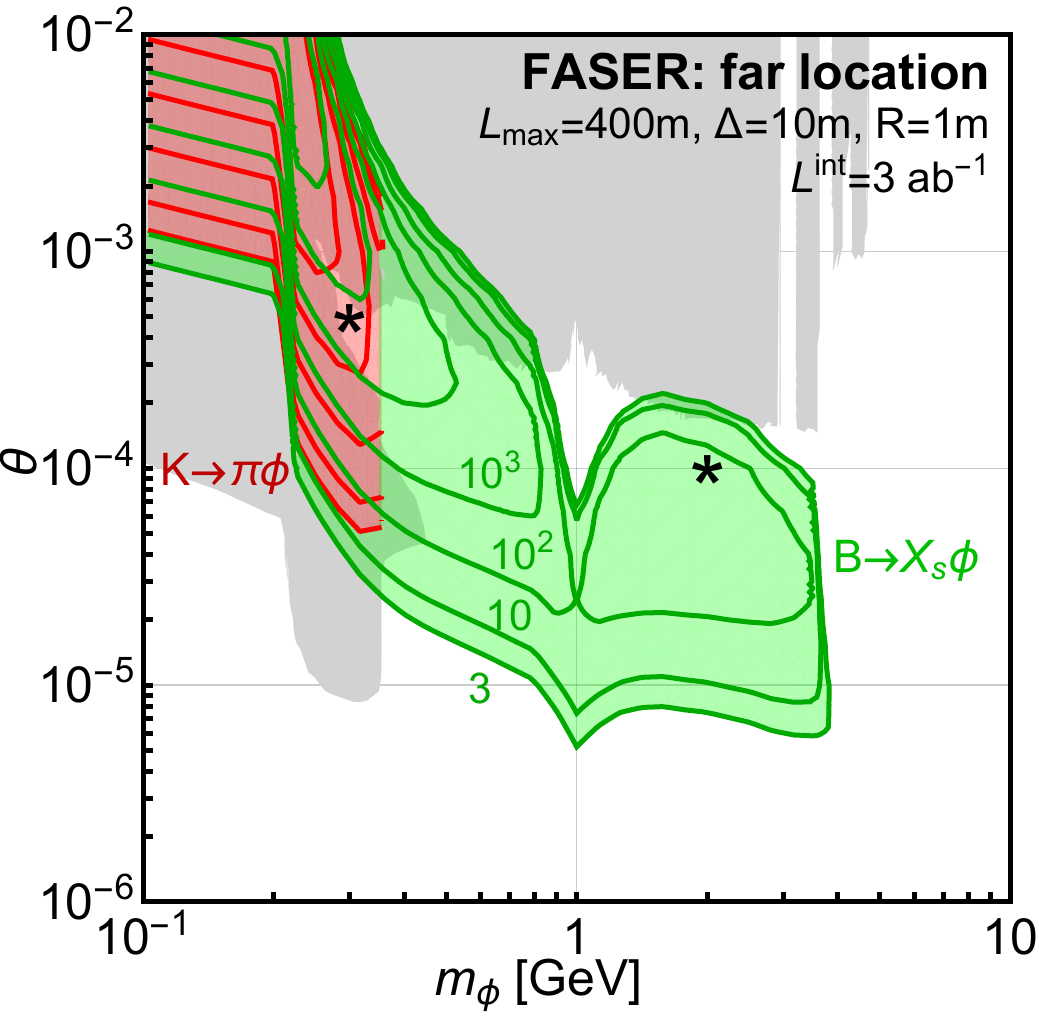} \\
\includegraphics[width=0.47\textwidth]{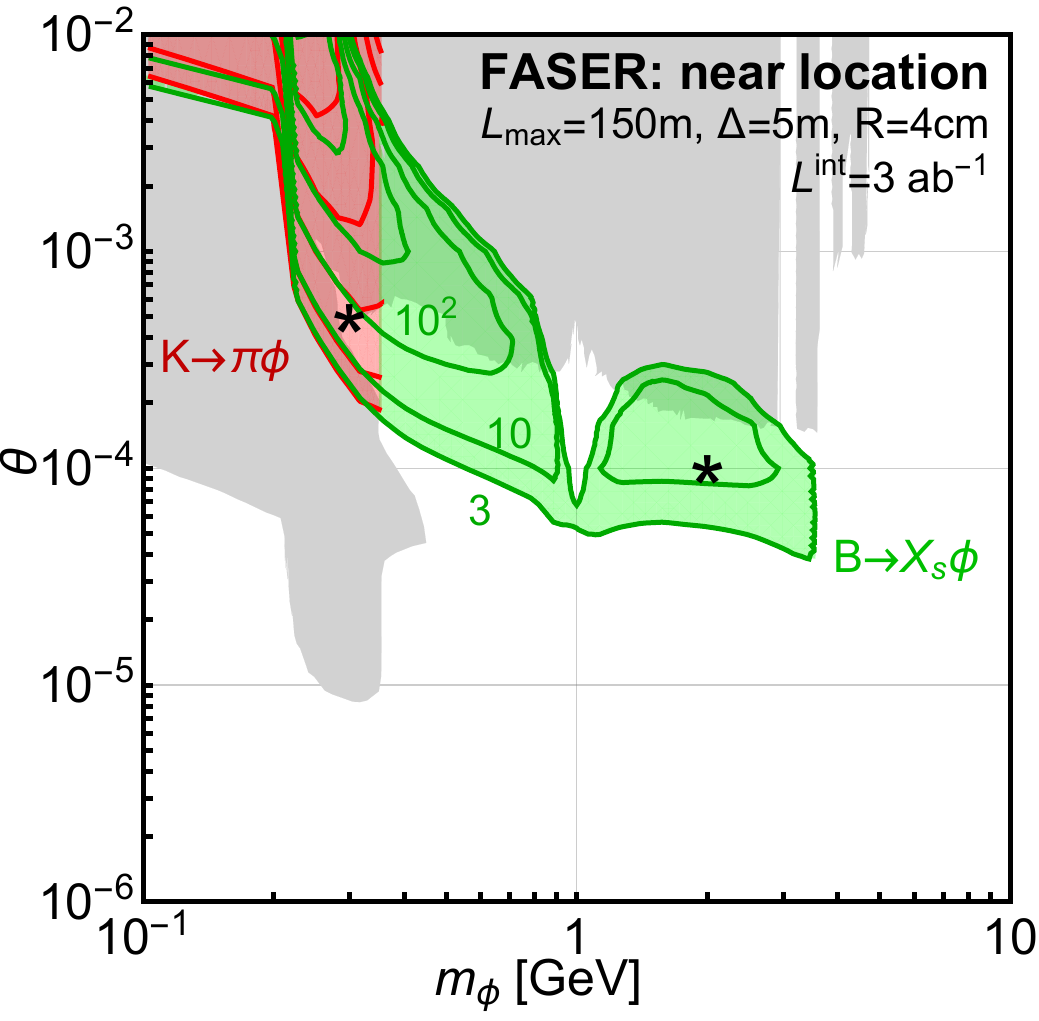} \quad
\includegraphics[width=0.47\textwidth]{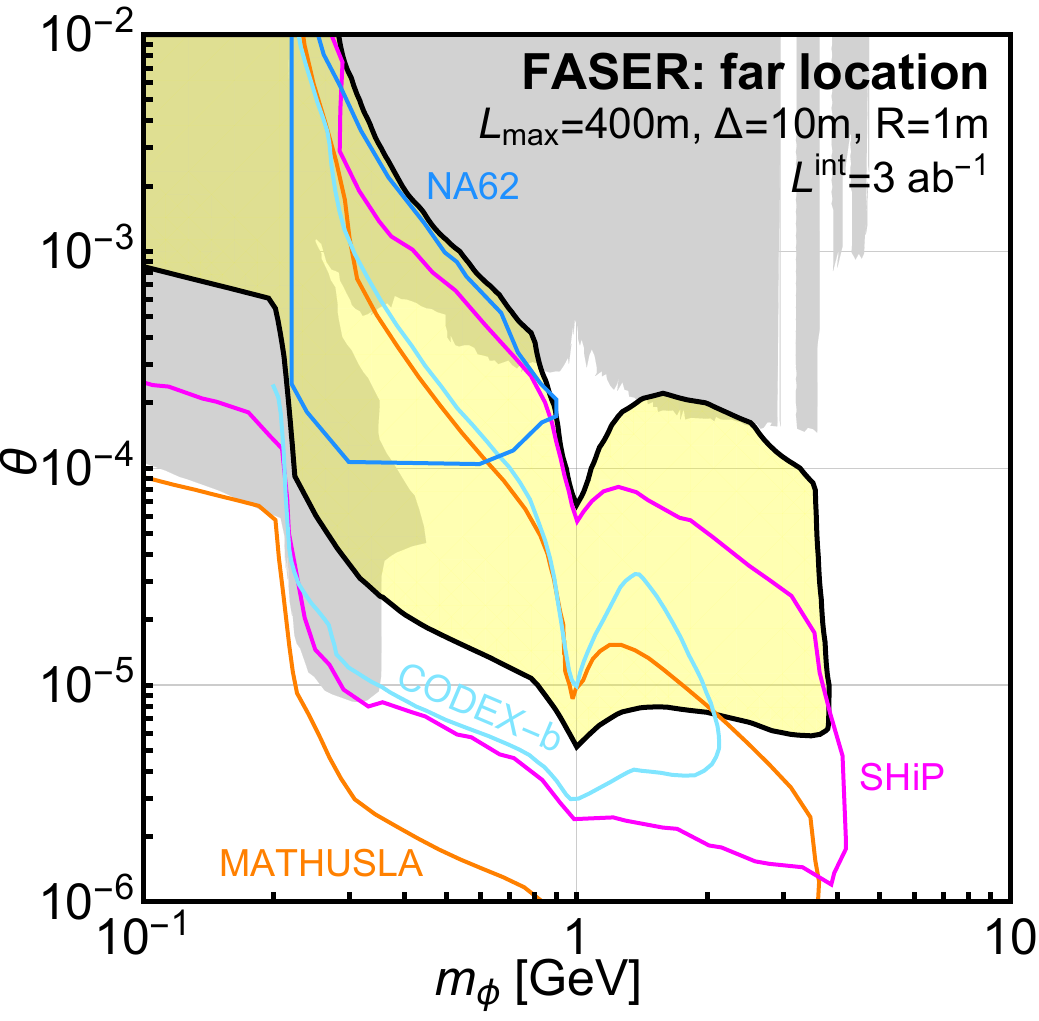}
\caption{Number of signal events $\nsig$ in dark Higgs parameter space for the far detector location with $R=20~\cm$ (top left) and $R=1~\m$ (top right) and for the near detector location (bottom left), given an integrated luminosity of $3~\iab$ at the 13 TeV LHC. As indicated, the contours are for $\nsig = 3, 10, 100, 1000, \ldots$ from the processes $b\to s \phi$ (green) and  $K  \to \pi \phi$ (red). The gray shaded regions are excluded by current experimental bounds. The black stars correspond to the representative parameter-space points discussed in the text. The bottom right panel shows the exclusion reach for FASER at the far location for  $R=1~\m$ (solid black line) along with the projected reaches of other proposed experiments that search for long-lived particles. 
}
\label{fig:eventrate-mixing}
\end{figure}

We now estimate the reach in dark Higgs parameter space at FASER.  In \figref{eventrate-mixing} we show contours of the expected number of signal events, assuming 100\% efficiency in detecting dark Higgs boson decays in FASER.  The green and red contours indicate the production processes $B \to X_s \phi$ and $K \to \pi \phi$, where all kaon species are included.  The three panels correspond to the three detector setups specified in \eqref{detector-location}. In our simulations, we employed a cut on the dark Higgs momentum, $p_\phi>100~\gev$, which is anyway effectively imposed by the requirement that the dark Higgs bosons propagate to the detector locations considered.

The gray-shaded regions of parameter space have already been excluded by previous experiments. The low-mass regime is excluded primarily by the CHARM experiment, a beam dump experiment searching for long-lived particles produced mostly in kaon decays and decaying into lepton or photon pairs~\cite{Bergsma:1985qz,Bezrukov:2009yw}. In the intermediate-mass regime, $2m_\mu < m_\phi < m_B-m_{K^{(*)}}$, the strongest constraints are from LHCb searches for $B^0 \to K^{*0} (\phi \to\mu^+ \mu^-)$~\cite{Aaij:2015tna} and $B^+ \to K^+ (\phi \to\mu^+ \mu^-)$~\cite{Aaij:2016qsm} containing a possibly displaced di-muon resonance. 

From \figref{eventrate-mixing}, we see that, if backgrounds can be reduced to negligible levels, as discussed in Ref.~\cite{Feng:2017uoz}, FASER will be able to probe the remaining gap between the CHARM and LHCb constraints using dark Higgs bosons produced in both the $K \to \pi \phi$ and $B \to X_s \phi$ channels.  The $B \to X_s \phi$ channel will further be able to probe a currently unconstrained region of parameter space in the mass range $2m_\mu \alt m_\phi \alt 2m_\tau $ for couplings $10^{-5} \alt \theta \alt 10^{-3}$. The reach at FASER drops rapidly when the $\phi \to \tau\tau, DD$ channels open up, given the corresponding sharp drop in dark Higgs lifetime. Comparing the three panels in \figref{eventrate-mixing}, we see that a far detector with relatively large radius $R = 1~\m$ has the largest reach.

In \cite{Feng:2017uoz}, we made a first attempt to estimate the backgrounds and considered a variety of SM processes that could mimic the signal. We focused on processes resulting in two high-energy, opposite-charge tracks that are simultaneous within the detector time resolution and to point back to the IP within the detector angular resolution. We found that the natural (rock) and LHC infrastructure (magnets and absorbers) shielding eliminates most potential background processes, but leaves the possibility for neutrino-induced and beam-induced backgrounds.

Neutrino-induced backgrounds can arise from high energy neutrinos produced close to the IP that propagate toward FASER and interact with the material inside the detector. The leading contribution in this case is from the process $\nu_\mu N\to\mu^-\pi^+X$, but due to the small nuclear recoil this process produces two charged tracks that are highly asymmetric in energy, in contrast to the signal.  For the far detector, we estimated the rate for this background and found it to be negligible. Other neutrino backgrounds may arise from kaon production in the rock before the detector and its subsequent decay into two or more charged tracks inside the detector. The production of high energy kaons is, however, suppressed from similar nuclear recoil arguments, and the kaon energy loss in material is substantial. We found these backgrounds to be negligible as well.	

Beam-induced backgrounds are substantially harder to estimate.
For the far location, which is well shielded from hadrons and electromagnetic radiation from the IP, muons produced in beam-gas interactions are likely the dominant background. 
As a first attempt to estimate such backgrounds, we considered the possibility that two uncorrelated muons produced in beam-gas interactions arrive at FASER within a time shorter than the detector time resolution.  Taking into account the production rate of high energy muons in beam-gas collisions~\cite{Aad:2013zwa}, the LHC bunch crossing structure and rate, and assuming a reasonable detector time resolution of $\delta t=100~\ps$, we found a negligible di-muon rate. A more realistic background analysis requires the exact material and design layout of the LHC tunnels, and models of radiation-matter interactions as embedded in tools like FLUKA~\cite{Ferrari:2005zk,FLUKA2014} and MARS~\cite{Mokhov:1995wa,Mokhov:2012ke}.  Such an analysis is currently underway, but is beyond the scope of this work.

Last, it is important to note that, if the background is non-negligible, but well-estimated, the resulting reach can be estimated from the event yield contours in \figref{eventrate-mixing}.  As seen there, even if 10 or 100 signal events are required, FASER can still probe significant regions of currently unexplored parameter space.

The projected reaches~\cite{Evans:2017lvd} of NA62~\cite{NA62} and the proposed SHiP~\cite{Alekhin:2015byh}, MATHUSLA~\cite{Chou:2016lxi}, and CODEX-b~\cite{Gligorov:2017nwh} experiments are shown in the bottom right panel of \figref{eventrate-mixing}.\footnote{SeaQuest, a $120~\gev$ proton beam dump experiment at Fermilab, is not competitive in the search for dark Higgs bosons because its center-of-mass energy results in a small $B$ meson production rate. This is in contrast to the case of dark photons, which are produced in light meson decays and through dark bremsstrahlung~\cite{Gardner:2015wea}.}  These experiments probe similar ranges of dark Higgs lifetimes, and are therefore only sensitive in the $2 m_{\mu} \alt \mphi \alt 2 m_{\tau}$ range, since for $\mphi > 2m_{\tau}$ ($\mphi < 2m_{\mu}$), the dark Higgs is too prompt (long-lived). FASER's reach exceeds the projections for NA62.  Its reach in $\theta$ is complementary to the three other experiments: while there is significant overlap, SHiP, MATHUSLA, and CODEX-b are more sensitive at relatively low $\theta$, while FASER covers the relatively high $\theta$ region, particularly the region with $\theta \sim 10^{-4}$ and $\mphi>1~\gev$, which is not covered by the other experiments.

To understand the complementarity of FASER and other experiments, as a specific example, consider FASER and SHiP with $\theta \sim 10^{-4}$. The advantage of a fixed-target experiment like SHiP is that it has far more $pp$ collisions than a collider experiment.  However, the number of $B$ mesons produced at SHiP, $7\times 10^{13}$~\cite{Alekhin:2015byh}, is less than the number produced at the LHC for FASER, $1.42 \times 10^{15}$, because SHiP's center-of-mass energy is much lower.  Even more important is the difference in the probabilities of dark Higgs bosons decaying in the detector (see \eqref{eq:P_decay_in_volume}).  In our notation, the distance scales of SHiP are $R^{\rm SHiP} \approx 2.5~\m$, and $\lmax^{\rm SHiP}=120~\m$~\cite{Alekhin:2015byh}.  From \figref{decays} we see that SHiP's beam energy is too small to produce dark Higgs bosons with the $\sim\tev$ momenta that are required for $\bar d \sim \lmax^{\rm SHiP}$. Instead, $B$ mesons produced at SHiP can have energies of at most $E_B = 400~\gev$ which leads to a suppression of the dark Higgs event rate of at least $\exp(-\tev/E_\phi)\approx 0.08$, where we have assumed the maximal possible energy $E_\phi = E_B =400~\gev$. In practice, most $B$ mesons are produced with far smaller energies, only part of which is transferred to the dark Higgs, implying a far stronger exponential suppression.  For example, the SHiP collaboration~\cite{Alekhin:2015byh} (following CHARM~\cite{Bergsma:1985qz}), finds a mean dark Higgs energy of $E_\phi \sim 25~\gev$, implying a suppression factor of $\exp(-\tev/E_\phi)\approx 4 \times 10^{-18}$.  Last, SHiP's angular acceptance requires $\theta_\phi \approx p_T/E_\phi < R^{\rm SHiP}/\lmax^{\rm SHiP} \approx 20~\mrad$, while for dark Higgs bosons with $E_{\phi} \sim 100~\gev$ and $p_T \sim m_B$, we find $\theta_\phi \sim m_B/E_\phi \sim 50~\mrad$.  As shown in \figref{radius}, at least for FASER on the far location with $R = 1~\m$, essentially all dark Higgs bosons that decay near $\lmax$ satisfy the angular acceptance requirement.   All of these effects lead to FASER having better coverage than SHiP for $\theta \sim 10^{-4}$.   Of course, for low values of $\theta \alt 10^{-5}$, the decay lengths are larger, and, as we see in \figref{eventrate-mixing}, SHiP is a more sensitive experiment.  FASER is sensitive to such small $\theta$ only through processes induced by the trilinear $h \phi \phi$ coupling, to which we now turn.

%%%%%%%%%%%%%%%%%%%%%%%%%%%%%
% Trilinear
%%%%%%%%%%%%%%%%%%%%%%%%%%%%%
\section{Probes of the Dark Higgs Boson Trilinear Coupling}
\label{sec:trilinear}

FASER can also probe the trilinear coupling $h \phi \phi$, as this coupling induces the double dark Higgs production process $b \to s h^* \to s \phi \phi$, leading to dark Higgs bosons that then decay into SM particles. The production rate is controlled by the coupling $\lambda$ in \eqref{eq:Lphysical}, while the lifetime depends on the mixing angle $\theta$. In \secref{pairproduction} we discuss the kinematic distributions of dark Higgs bosons produced through $b \to s \phi \phi$.  In \secref{trilinearreach}, we then determine the number of dark Higgs bosons from double dark Higgs events that could be seen in FASER.

\subsection{Dark Higgs Pair Production in $B$ Decays}
\label{sec:pairproduction}

The $b\to s\phi\phi$ transition may be described by the effective Lagrangian~\cite{Bird:2004ts}
\be
\mathcal{L}_{bs\phi\phi}=\frac{1}{2}  m_b ( C \bar{s}_L b_R + C^* \bar{b}_L s_R )\phi^2 \ ,
\ee
where
\be
C =\frac{3 \lambda }{8 \pi^2  v^2 }  \frac{m_t^2}{m_h^2} V^*_{ts} V_{tb} = 4.9 \times 10^{-8}~\gev^{-2} \cdot \lambda \ .
\ee
Following Ref.~\cite{Altmannshofer:2009ma}, the inclusive differential decay width is
\be
\frac{d\Gamma_{b\to s\phi\phi}}{dq^2} 
= \frac{C^2}{256 \pi^3 m_b} \left( 1-\frac{4 \mphi^2}{q^2} \right)^{1/2} \lambda^{1/2}(m_b^2, m_s^2, q^2)
\left[ (m_b-  m_s)^2 -q^2 \right] ,
\ee
where $q^2=(p_b-p_s)^2$. Taking the limit $m_s \to 0$ and integrating over $q^2$ from $4\mphi^2$ to $m_b^2$, we find
\be
B(b\to s\phi\phi) = \frac{\Gamma_{b\to s\phi\phi}}{\Gamma_B} 
= \frac{1}{\Gamma_B} \ \frac{C^2 m_b^5 }{256 \pi^3 } \  f \! \left( \frac{\mphi}{m_b}\right) \ ,
\ee
where
\be
f(x) = \frac{1}{3}  \sqrt{1-4x^2} (1 + 5x^2 - 6 x^4) -4 x^2 (1 - 2 x^2 + 2 x^4 ) \log\left[  \frac{1}{2x} \left(1 +\sqrt{1-4 x^2}\right) \right] .
\ee
For $\mphi = 1~\gev$, $B(b\to s\phi\phi) \simeq 2.1 \times 10^{-4} \cdot \lambda^2$.

The decay kinematics can be specified by five parameters: $q^2$, the polar and azimuthal angles of the off-shell Higgs in the $b$-quark rest frame, and the polar and azimuthal angles of the dark Higgs bosons in the off-shell Higgs rest frame.  To simulate the 3-body decay, we scan over $q^2$ and integrate over the rest of the parameters via Monte Carlo.

\begin{figure}[tb]
\centering
\includegraphics[width=0.47\textwidth]{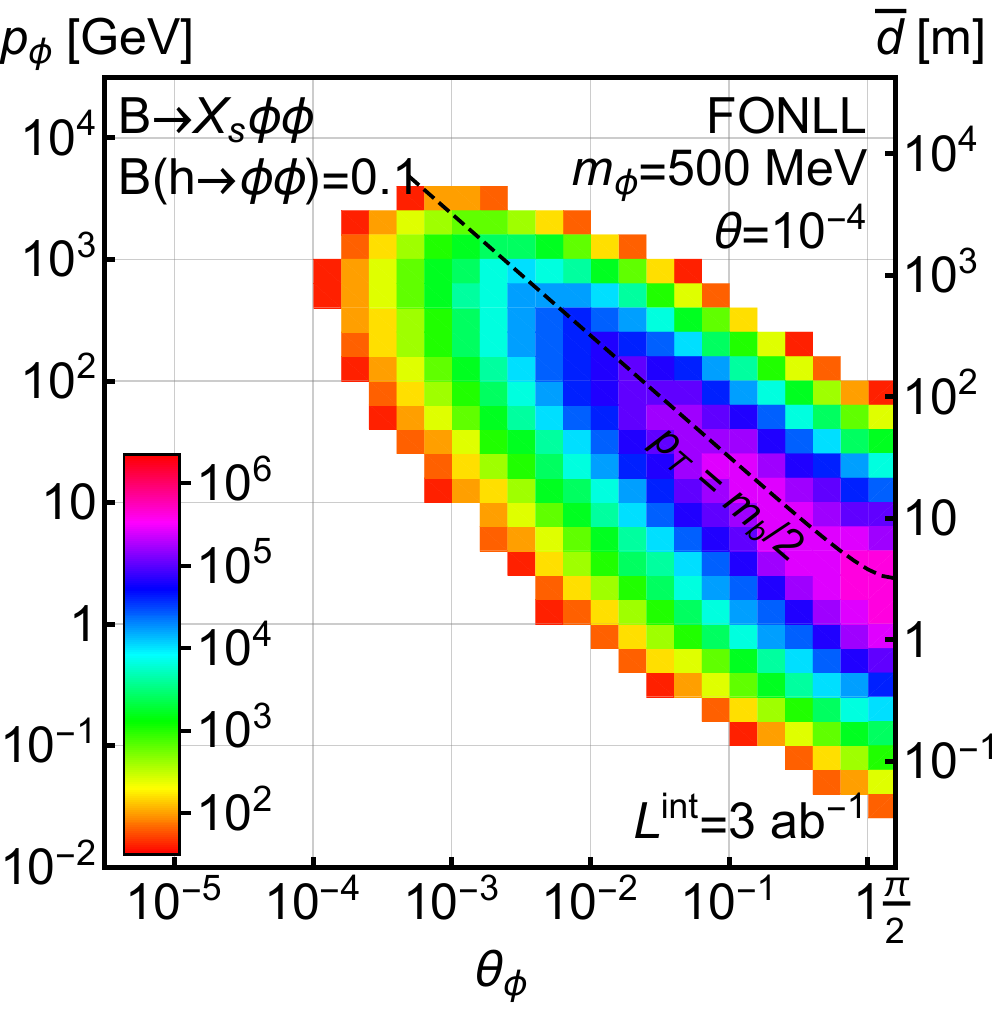} \quad
\includegraphics[width=0.47\textwidth]{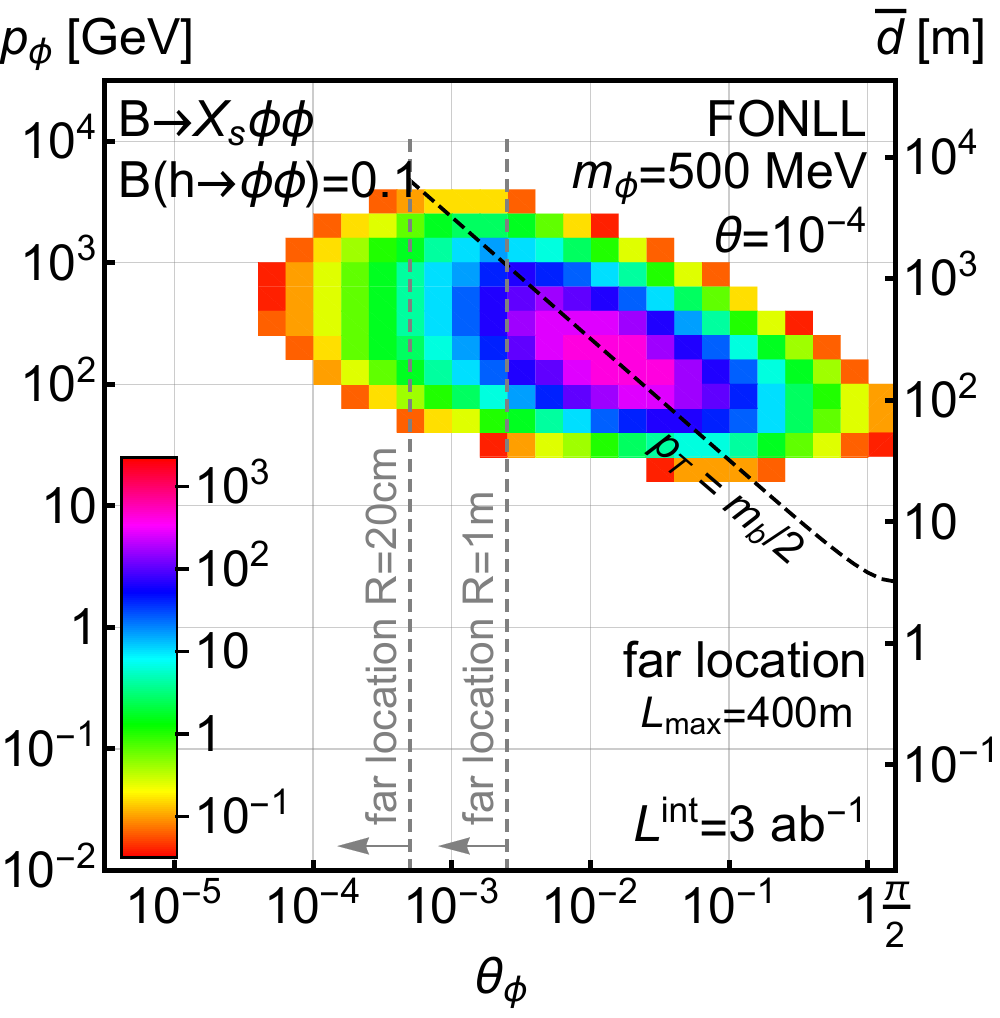} \\
\includegraphics[width=0.47\textwidth]{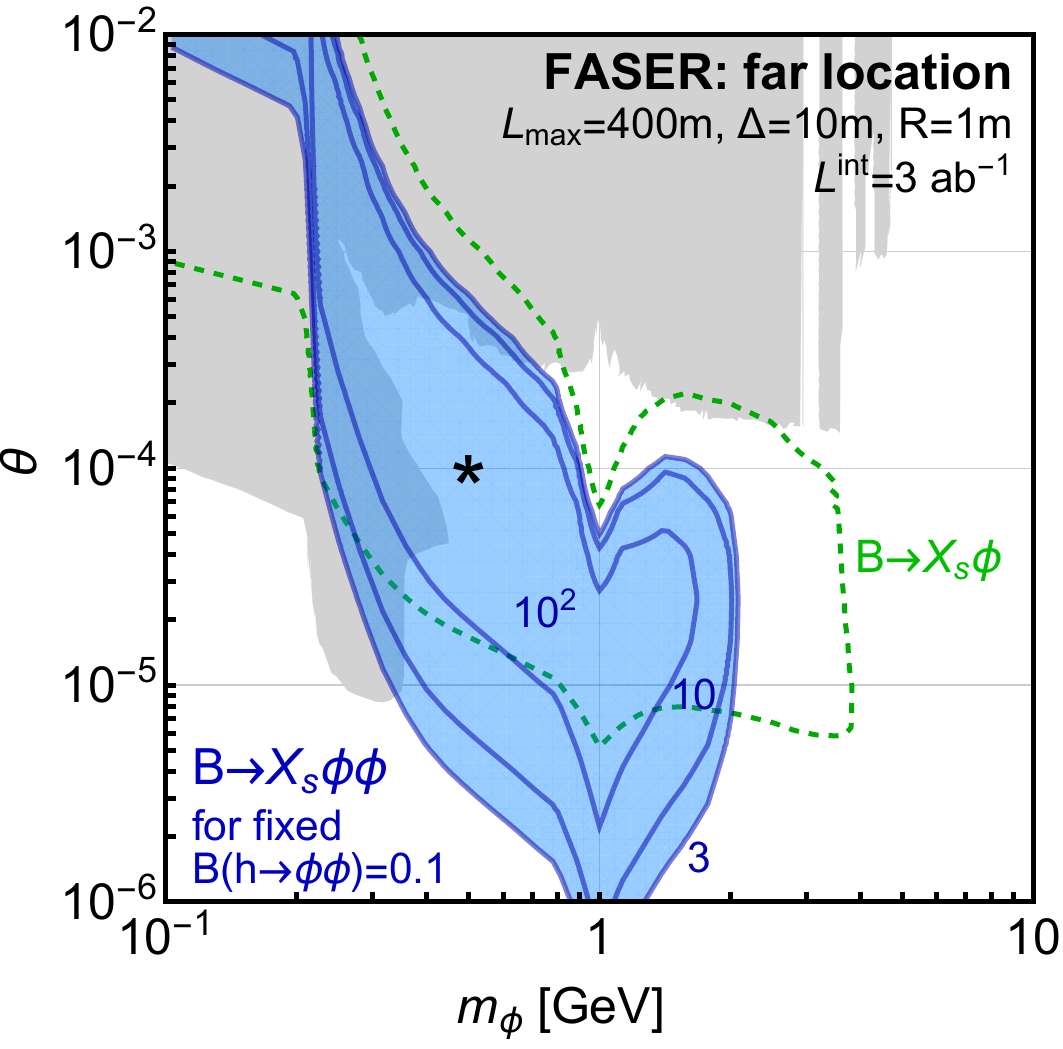}
\caption{Top left: Distribution of dark Higgs bosons $\phi$ in the $(\theta_\phi, p_\phi)$ plane from the process $b \to s \phi\phi$, for $(\mphi, \theta)=(500~\mev, 10^{-4})$ and $\lambda = 4.6 \times 10^{-3}$, corresponding to $B(h\to\phi\phi)=0.1$. The results are for the 13 TeV LHC with integrated luminosity $L=3~\iab$. Top right: Same distribution, but for dark Higgs bosons that decay within the range $(\lmin,\lmax) = (390~\m, 400~\m)$. Bottom: Number of signal events $\nsig= 3, 10, 100$ from the process $b \to s \phi\phi$ in the dark Higgs parameter plane $(\mphi, \theta)$, assuming $\lambda = 4.6 \times 10^{-3}$, and FASER at the far location with $R=1~\m$. The star indicates the representative parameter space point $(\mphi, \theta)=(500~\mev, 10^{-4})$. }
\label{fig:DarkHiggsPair}
\end{figure}

In the top left panel of \figref{DarkHiggsPair}, we show the distribution of dark Higgs bosons produced in $b \to s \phi \phi$ in the $(\theta_\phi, p_\phi)$ plane.  We have set $m_\phi=500~\mev$ and $\lambda = 4.6 \times 10^{-3}$, corresponding to $B(h\to\phi\phi)=0.1$ (see below). The typical transverse momentum of the produced dark Higgs bosons is $p_T \sim m_b / 2$. The top right panel of \figref{DarkHiggsPair} shows the distribution of dark Higgs bosons that decay in the far detector range $(\lmin,\lmax) = (390~\m, 400~\m)$, assuming a coupling $\theta=10^{-4}$ for illustration. For these parameters, we see that, despite the highly suppressed branching fraction for $b \to s \phi \phi$, hundreds of dark Higgs bosons can be produced and decay in FASER.

\subsection{Trilinear Coupling Reach}
\label{sec:trilinearreach}

We now evaluate FASER's sensitivity to $\lambda$ and compare it to other probes. In particular, the trilinear $h\phi\phi$ coupling also induces the SM Higgs decay $h \to \phi\phi$ with branching fraction
\be
B(h\to \phi\phi) 
= \frac{\Gamma(h\to \phi\phi)}{\Gamma_h} 
\approx \frac{1}{\Gamma_h^{\text{SM}}} \,
\frac{ \lambda^2 v^2}{8 \pi m_h} \left(1-\frac{4 \mphi^2}{m_h^2} \right)^{1/2} 
\simeq 4700 \cdot  \lambda^2 \ . 
\ee
Current limits depend on both $\lambda$ and $\theta$, but for small $\theta$, where the $\phi$ lifetime is large, these events lead to invisible Higgs decays, which are constrained by searches at CMS~\cite{Khachatryan:2016whc} and ATLAS~\cite{Aad:2015txa,Aaboud:2017bja}.  The most stringent current bound of $B(h \to \text{inv}) < 0.24$ implies $\lambda< 7.1 \times 10^{-3}$.

In the bottom panel of \figref{DarkHiggsPair}, we set $\lambda = 4.6 \times 10^{-3}$, corresponding to $B(h \to \phi \phi) = 0.1$, roughly the sensitivity to invisible decays of the LHC with $300~\ifb$. We then show the number of dark Higgs bosons from the pair production process that decay in FASER at the far location with $R = 1~\m$. Despite the highly suppressed pair production process, hundreds of dark Higgs bosons from this process could be detected by FASER.  The region of parameter space probed by the single and pair production processes are complementary: for currently viable values of $\lambda$, there are regions of the $(m_\phi, \theta)$ parameter space that can produce pair production signals without single production signals, and vice versa.   In this way, FASER is similar to MATHUSLA, which is sensitive to dark Higgs bosons from meson decay and also from pair production in $h \to \phi \phi$~\cite{Evans:2017lvd}.  We note, however, that, because the regions of sensitivity to single and double production have significant overlap, if a signal is seen, more detailed work is required to determine its origin or bound particular sources.

%%%%%%%%%%%%%%%%%%%%%%%%%%%%%
% Cosmology
%%%%%%%%%%%%%%%%%%%%%%%%%%%%%
\section{Cosmological Connections}
\label{sec:cosmology}

As mentioned in \secref{introduction}, searches for dark Higgs bosons have implications for DM and inflation. 

\subsection{Dark Matter}

Searches for dark Higgs bosons have implications for dark matter if they are mediators between the SM and dark sectors.  As an example, suppose the dark Higgs boson couples to the SM as given in \eqref{eq:Lphysical} and also to Majorana fermion DM $X$ through the interaction (see, \eg,~\cite{Krnjaic:2015mbs}, and recent discussion of complex scalar and pseudo-Dirac cases in Ref.~\cite{Darme:2017glc})
\be
\mathcal{L} \supset -\frac{1}{2}\kappa\,\phi\,\bar{X}X \ .
\ee
We will assume $m_{X} > m_\phi$, and that the $X$ thermal relic density is determined by the annihilation cross section $\sigma (X X \to \phi \phi) \sim \kappa^4$, since annihilation to SM final states is suppressed by powers of $\theta$. 

For given values of $m_\phi$ and $m_X$, the thermal relic density determines $\kappa$, and bounds on the direct detection scattering cross section constrain $\sigma (X N \to X N) \sim \kappa^2 \theta^2$.  As a result, for fixed values of the ratio $\mphi/m_X$, current direct detection limits~\cite{Angloher:2015ewa,Agnese:2015nto,Akerib:2016vxi,Amole:2017dex,Aprile:2017iyp,Cui:2017nnn} constrain the $(\mphi,\theta)$ plane.  These constraints are shown in  \figref{DM} for various values of $\mphi/m_X$, along with the projected reach of future direct detection experiments~\cite{Battaglieri:2017aum}. We see that, depending on $\mphi/m_X$, FASER can probe regions of the parameter space that are beyond any proposed direct detection experiment.  Of course, if signals are seen in both FASER and direct detection, they will provide complementary probes of the dark sector. 

\begin{figure}[tb]
\centering
\includegraphics[width=0.47\textwidth]{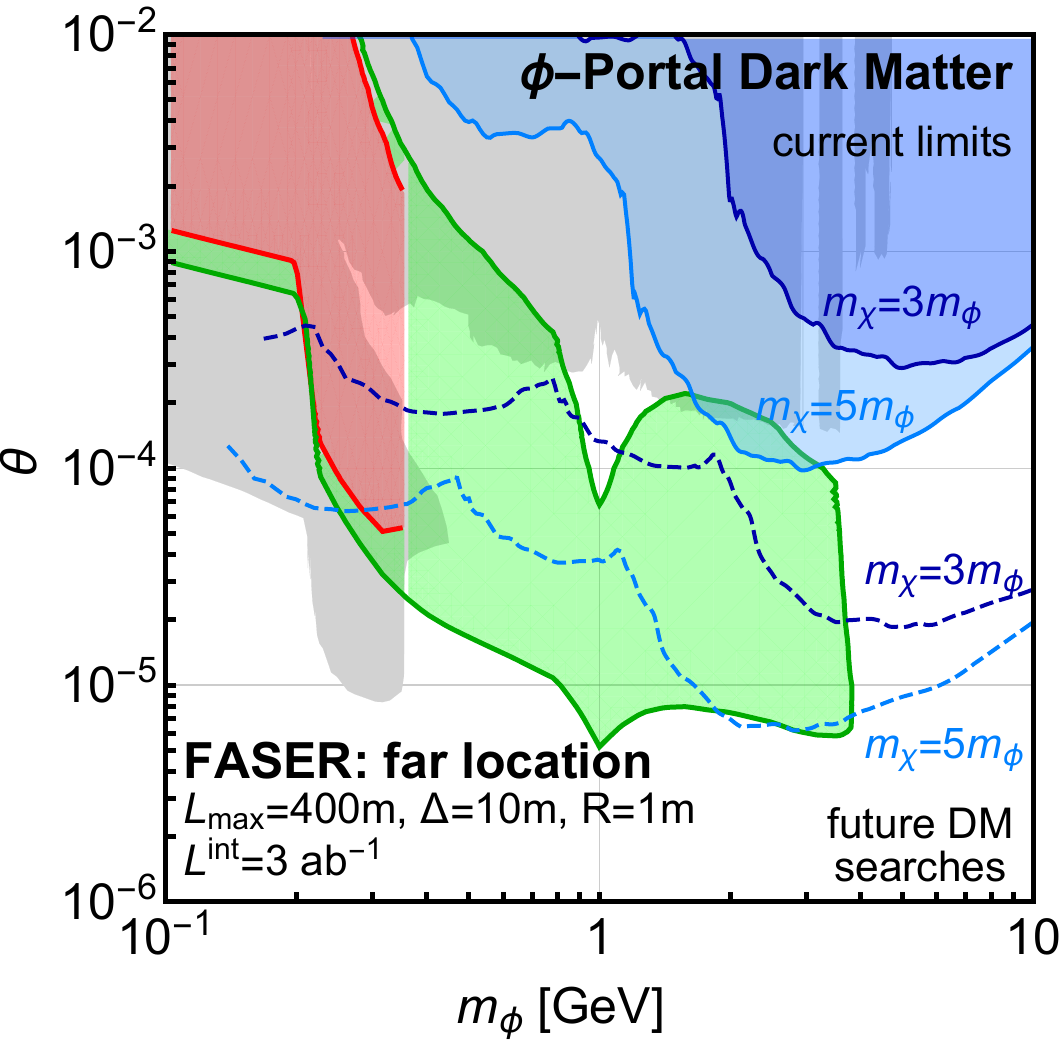}
\caption{Regions of dark Higgs parameter space that are probed by direct detection searches for dark matter, assuming the dark Higgs boson mediates interactions between the SM and a Majorana fermion thermal relic $X$.  The gray shaded regions are excluded by current constraints.  The blue shaded regions are excluded by current direct detection bounds~\cite{Angloher:2015ewa,Agnese:2015nto,Akerib:2016vxi,Amole:2017dex,Aprile:2017iyp,Cui:2017nnn} for $m_X = 3 m_\phi$ and $5 m_\phi$.  The dashed contours represent projected sensitivities of future direct detection experiments~\cite{Battaglieri:2017aum}.  As in \figref{eventrate-mixing}, the green and red regions are the reach of FASER from $B$ and $K$ decays, respectively.}
\label{fig:DM}
\end{figure}

\subsection{Inflation}

Inflatons that are light and can be produced in particle physics experiments are phenomenologically an appealing alternative to the standard paradigm, in which inflatons are assumed to be heavy and therefore effectively decoupled from the SM. In some of these models, the light inflaton is identified with the dark Higgs boson~\cite{Shaposhnikov:2006xi,Bezrukov:2009yw,Bezrukov:2013fca,Bramante:2016yju}. The scalar Lagrangian is then typically that of \eqref{Lagrangian} with some parameters set to zero for simplicity. 

Specifically, in the model discussed in Refs.~\cite{Shaposhnikov:2006xi,Bezrukov:2009yw,Bezrukov:2013fca}, one sets $\mu=\mu'_{12}=\mu'_3=0$ in \eqref{Lagrangian}.  Electroweak symmetry breaking is driven by a non-zero inflaton vev in the Higgs-inflaton mixing term. It was originally found that the preferred mass range for the inflaton in this model is $270~\mev \lesssim m_\phi \lesssim 1.8~\gev$~\cite{Bezrukov:2009yw}, with a range of $\theta$ that could be probed by FASER. Since the early analyses, however, the SM Higgs boson discovery~\cite{Aad:2012tfa,Chatrchyan:2012xdj} and more recent cosmological data~\cite{Ade:2015xua} overconstrain the original model.  To alleviate the tension between the measured and predicted values of the tensor-to-scalar ratio $r$, one can, \eg, introduce a non-minimal coupling of the inflaton field to gravity, as discussed in Ref.~\cite{Bezrukov:2013fca}. The relatively large values of $r$ inferred from such a model can be probed by future CMB observations; for a review, see~\cite{Abazajian:2016yjj}. At the same time, a complementary search for the light inflaton in FASER would allow one to thoroughly investigate the consistency of the model with the experimental and observational data.

Alternatively, in models with low-scale inflation that predict very small values of $r$ beyond the reach of CMB searches, the search for a light inflaton in FASER could be a rare opportunity to test inflation experimentally. In particular, in Ref.~\cite{Bramante:2016yju} such a possibility is discussed for a model with quartic hilltop inflation. Interestingly, these type of models can also contain another dark-Higgs-like scalar that can be probed in FASER, namely, the curvaton, which is introduced to reproduce the observed spectrum of CMB perturbations.

%%%%%%%%%%%%%%%%%%%%%%%%%%%%%
% Conclusions 
%%%%%%%%%%%%%%%%%%%%%%%%%%%%%
\section{Conclusions}
\label{sec:conclusions}

Accessible new particles can be either relatively strongly interacting and heavy, or very weakly interacting and light.  In the latter case, the extraordinary event rates at the LHC, especially in the upcoming high luminosity era, mean that even extremely weakly-interacting new particles can be discovered in the low $p_T$ region along the beamline downstream of the ATLAS and CMS IPs.  The motivation of FASER, ForwArd Search ExpeRiment at the LHC, is to exploit this opportunity to discover new physics with a small, inexpensive detector.  

In this study, we have explored the potential for FASER to discover dark Higgs bosons, hidden scalars that couple to the SM through the renormalizable Higgs portal interaction.  As with SM Higgs bosons, dark Higgs bosons couple preferentially to heavy SM fermions, and so they are dominantly produced through the processes $B \to X_s \phi$ and $K \to \pi \phi$.  At the 13 TeV LHC with $3~\iab$, FASER will greatly extend the discovery prospects for dark Higgs bosons.  As many as $10^4$ dark Higgs boson can be seen in FASER in currently unexplored parameter space, and FASER may discover dark Higgs bosons with masses $\mphi = 0.2 - 3.5~\gev$ and mixing angles $\theta \sim 10^{-5} - 10^{-3}$.  Although FASER's sensitivity given in \figref{eventrate-mixing} clearly shows a significant overlap with other proposed experiments, FASER is uniquely sensitive in the parameter space with $\mphi \agt 1~\gev$ and $\theta\sim 10^{-4}$, for which dark Higgs bosons are typically too prompt to be detected by the other experiments. This result is due to the combination of FASER's geometric acceptance with the $(\theta_\phi, p_\phi)$ kinematic distribution of dark Higgses produced at the LHC. In addition, FASER may also discover dark Higgs bosons produced through $b \to s h^* \to s \phi \phi$, which probes the trilinear coupling $h \phi \phi$ and is complementary to probes of the rare SM Higgs decay $h \to \phi \phi$; the reach in this case is shown in \figref{DarkHiggsPair}.

FASER will probe models of cosmological interest.  For example, dark Higgs bosons may mediate interactions between the SM and dark matter particles $X$.  For $m_\phi , m_X \sim \gev$, the required couplings for a thermal relic are $\theta \sim 10^{-5} - 10^{-3}$, depending on the hidden sector coupling. Such couplings are exactly in FASER's range of sensitivity.  We have shown that FASER's sensitivity also surpasses current direct detection searches.  In addition, FASER is sensitive to viable regions of parameter space in scenarios in which the dark Higgs boson is also the inflaton.  If a signal is seen at FASER, it will shed light on inflation in these models, a rare opportunity to probe inflation in particle physics experiments.

The search for dark Higgs bosons has many interesting features when compared to the search for dark photons at FASER~\cite{Feng:2017uoz}.  These two possibilities probe renormalizable couplings of the SM to the hidden sector, which can reasonably be expected to be the leading couplings in many cases.  At the same time, there are several qualitative differences:
\begin{itemize}[itemsep=0.03cm,topsep=0.15cm,leftmargin=1.5em]
\item Dark photons are dominantly produced with $p_T \sim \Lambda_{\text{QCD}}$, and so are highly collimated even 400 m from the IP, where they may be collected with a cylindrical detector with a radius of only $R = 20~\cm$.  In contrast, dark Higgs bosons are predominantly produced in $B$ decays with $p_T \sim m_B$, and they are therefore less collimated.  A detector with radius 1 m can improve search prospects significantly.  
\item FASER is mainly sensitive to the parameter space in which the dark photons are relatively short-lived (large $\epsilon$), and arrive in FASER only due to their high energy, while the long-lived regime (small $\epsilon$) is already mostly excluded. In contrast, in the dark Higgs case, one typically probes relatively long lifetimes, which, in turn, require smaller boosts and again imply larger angles with respect to the beam axis. 
\item As shown in \figref{bf}, the dark Higgs predominantly decays into the heaviest allowed decay mode. At FASER, these are mainly muon, pion and kaon pairs, depending on the dark Higgs mass. A tracker-based technology in combination with a magnetic field will detect the charged particle final states.  To detect neutral decay modes, such as $\phi \to \pi^0 \pi^0$, these components could be augmented by an electromagnetic calorimeter and, possibly, a ring-Cherenkov detector.
\end{itemize}

Here and in Ref.~\cite{Feng:2017uoz}, we have considered dark Higgs bosons and dark photons separately.  Of course, these two new particles may naturally appear together in theories where the dark photon mass is generated by a non-zero vev of a dark Higgs boson.  (For a review see, \eg, Ref.~\cite{Raggi:2015yfk}.) In such theories, the dark photon and dark Higgs boson naturally have similar masses, and they may be simultaneously probed by FASER.  In addition, the impact of the interaction term between dark Higgs boson and dark photons, $\phi  A'_\mu A'^\mu$, could significantly alter our discussion of the sensitivity reach of FASER. Specifically, dark Higgs boson decays into two dark photons, $\phi \to A'A'$, if kinematically allowed, could make a dark Higgs boson discovery much more challenging, but at the same time significantly improve the prospects for a dark photon detection. On the other hand, the number of dark Higgs bosons going towards FASER could be increased by additional processes in which dark photons are produced and then radiate off dark Higgs bosons.

%%%%%%%%%%%%%%%%%%%%%%%%%%%%%
% ACKNOWLEDGMENTS
%%%%%%%%%%%%%%%%%%%%%%%%%%%%%
\acknowledgments

We thank Joe Bramante and Jared Evans for useful discussions. This work is supported in part by NSF Grant No.~PHY-1620638. J.L.F. is supported in part by Simons Investigator Award \#376204.  I.G. is supported in part by DOE grant DOE-SC0010008. F.K. performed part of this work at the Aspen Center for Physics, which is supported by NSF Grant No.~PHY-1607611. S.T. is supported in part by the Polish Ministry of Science and Higher Education under research grant 1309/MOB/IV/2015/0.

\bibliography{darkhiggs}

\providecommand{\href}[2]{#2}\begingroup\raggedright\begin{thebibliography}{10}

\bibitem{Battaglieri:2017aum}
M.~Battaglieri {\em et al.}, ``{US Cosmic Visions: New Ideas in Dark Matter
  2017: Community Report},''
\href{http://arxiv.org/abs/1707.04591}{{\ttfamily arXiv:1707.04591 [hep-ph]}}.
%%CITATION = ARXIV:1707.04591;%%.

\bibitem{Feng:2017uoz}
J.~Feng, I.~Galon, F.~Kling, and S.~Trojanowski, ``{FASER: ForwArd Search
  ExpeRiment at the LHC},''
\href{http://arxiv.org/abs/1708.09389}{{\ttfamily arXiv:1708.09389 [hep-ph]}}.
%%CITATION = ARXIV:1708.09389;%%.

\bibitem{Patt:2006fw}
B.~Patt and F.~Wilczek, ``{Higgs-field portal into hidden sectors},''
\href{http://arxiv.org/abs/hep-ph/0605188}{{\ttfamily arXiv:hep-ph/0605188
  [hep-ph]}}.
%%CITATION = HEP-PH/0605188;%%.

\bibitem{Feng:2008ya}
J.~L. Feng and J.~Kumar, ``{The WIMPless Miracle: Dark-Matter Particles without
  Weak-Scale Masses or Weak Interactions},''
  \href{http://dx.doi.org/10.1103/PhysRevLett.101.231301}{{\em Phys. Rev.
  Lett.} {\bfseries 101} (2008) 231301},
\href{http://arxiv.org/abs/0803.4196}{{\ttfamily arXiv:0803.4196 [hep-ph]}}.
%%CITATION = ARXIV:0803.4196;%%.

\bibitem{Tulin:2017ara}
S.~Tulin and H.-B. Yu, ``{Dark Matter Self-interactions and Small Scale
  Structure},''
\href{http://arxiv.org/abs/1705.02358}{{\ttfamily arXiv:1705.02358 [hep-ph]}}.
%%CITATION = ARXIV:1705.02358;%%.

\bibitem{Shaposhnikov:2006xi}
M.~Shaposhnikov and I.~Tkachev, ``{The nuMSM, inflation, and dark matter},''
  \href{http://dx.doi.org/10.1016/j.physletb.2006.06.063}{{\em Phys. Lett.}
  {\bfseries B639} (2006) 414--417},
\href{http://arxiv.org/abs/hep-ph/0604236}{{\ttfamily arXiv:hep-ph/0604236
  [hep-ph]}}.
%%CITATION = HEP-PH/0604236;%%.

\bibitem{Bezrukov:2009yw}
F.~Bezrukov and D.~Gorbunov, ``{Light Inflaton Hunter's Guide},''
  \href{http://dx.doi.org/10.1007/JHEP05(2010)010}{{\em JHEP} {\bfseries 05}
  (2010) 010},
\href{http://arxiv.org/abs/0912.0390}{{\ttfamily arXiv:0912.0390 [hep-ph]}}.
%%CITATION = ARXIV:0912.0390;%%.

\bibitem{Bezrukov:2013fca}
F.~Bezrukov and D.~Gorbunov, ``{Light inflaton after LHC8 and WMAP9 results},''
  \href{http://dx.doi.org/10.1007/JHEP07(2013)140}{{\em JHEP} {\bfseries 07}
  (2013) 140},
\href{http://arxiv.org/abs/1303.4395}{{\ttfamily arXiv:1303.4395 [hep-ph]}}.
%%CITATION = ARXIV:1303.4395;%%.

\bibitem{Bramante:2016yju}
J.~Bramante, J.~Cook, A.~Delgado, and A.~Martin, ``{Low Scale Inflation at High
  Energy Colliders and Meson Factories},''
  \href{http://dx.doi.org/10.1103/PhysRevD.94.115012}{{\em Phys. Rev.}
  {\bfseries D94} (2016) 115012},
\href{http://arxiv.org/abs/1608.08625}{{\ttfamily arXiv:1608.08625 [hep-ph]}}.
%%CITATION = ARXIV:1608.08625;%%.

\bibitem{NA62}
{\bfseries NA62} Collaboration, G.~Lanfranchi, ``{Search for Hidden Sector
  Particles at NA62},''.
  \url{https://indico.cern.ch/event/466934/contributions/2583562/attachments/1489427/2314489/EPS_2017_Lanfranchi.pdf}.

\bibitem{Alekhin:2015byh}
S.~Alekhin {\em et al.}, ``{A facility to Search for Hidden Particles at the
  CERN SPS: the SHiP physics case},''
  \href{http://dx.doi.org/10.1088/0034-4885/79/12/124201}{{\em Rept. Prog.
  Phys.} {\bfseries 79} (2016) 124201},
\href{http://arxiv.org/abs/1504.04855}{{\ttfamily arXiv:1504.04855 [hep-ph]}}.
%%CITATION = ARXIV:1504.04855;%%.

\bibitem{Chou:2016lxi}
J.~P. Chou, D.~Curtin, and H.~J. Lubatti, ``{New Detectors to Explore the
  Lifetime Frontier},''
  \href{http://dx.doi.org/10.1016/j.physletb.2017.01.043}{{\em Phys. Lett.}
  {\bfseries B767} (2017) 29--36},
\href{http://arxiv.org/abs/1606.06298}{{\ttfamily arXiv:1606.06298 [hep-ph]}}.
%%CITATION = ARXIV:1606.06298;%%.

\bibitem{Curtin:2017izq}
D.~Curtin and M.~E. Peskin, ``{Analysis of Long Lived Particle Decays with the
  MATHUSLA Detector},''
\href{http://arxiv.org/abs/1705.06327}{{\ttfamily arXiv:1705.06327 [hep-ph]}}.
%%CITATION = ARXIV:1705.06327;%%.

\bibitem{Evans:2017lvd}
J.~A. Evans, ``{Detecting Hidden Particles with MATHUSLA},''
\href{http://arxiv.org/abs/1708.08503}{{\ttfamily arXiv:1708.08503 [hep-ph]}}.
%%CITATION = ARXIV:1708.08503;%%.

\bibitem{Gligorov:2017nwh}
V.~V. Gligorov, S.~Knapen, M.~Papucci, and D.~J. Robinson, ``{Searching for
  Long-lived Particles: A Compact Detector for Exotics at LHCb},''
\href{http://arxiv.org/abs/1708.09395}{{\ttfamily arXiv:1708.09395 [hep-ph]}}.
%%CITATION = ARXIV:1708.09395;%%.

\bibitem{Clarke:2013aya}
J.~D. Clarke, R.~Foot, and R.~R. Volkas, ``{Phenomenology of a very light
  scalar (100 MeV $< m_h <$ 10 GeV) mixing with the SM Higgs},''
  \href{http://dx.doi.org/10.1007/JHEP02(2014)123}{{\em JHEP} {\bfseries 02}
  (2014) 123},
\href{http://arxiv.org/abs/1310.8042}{{\ttfamily arXiv:1310.8042 [hep-ph]}}.
%%CITATION = ARXIV:1310.8042;%%.

\bibitem{Donoghue:1990xh}
J.~F. Donoghue, J.~Gasser, and H.~Leutwyler, ``{The Decay of a Light Higgs
  Boson},''
\href{http://dx.doi.org/10.1016/0550-3213(90)90474-R}{{\em Nucl. Phys.}
  {\bfseries B343} (1990) 341--368}.
%%CITATION = NUPHA,B343,341;%%.

\bibitem{Gunion:1989we}
J.~F. Gunion, H.~E. Haber, G.~L. Kane, and S.~Dawson, ``{The Higgs Hunter's
  Guide},''
{\em Front. Phys.} {\bfseries 80} (2000) 1--404.
%%CITATION = FRPHA,80,1;%%.

\bibitem{McKeen:2008gd}
D.~McKeen, ``{Constraining Light Bosons with Radiative $\Upsilon$(1S)
  Decays},'' \href{http://dx.doi.org/10.1103/PhysRevD.79.015007}{{\em Phys.
  Rev.} {\bfseries D79} (2009) 015007},
\href{http://arxiv.org/abs/0809.4787}{{\ttfamily arXiv:0809.4787 [hep-ph]}}.
%%CITATION = ARXIV:0809.4787;%%.

\bibitem{Grinstein:1988yu}
B.~Grinstein, L.~J. Hall, and L.~Randall, ``{Do B meson decays exclude a light
  Higgs?},''
\href{http://dx.doi.org/10.1016/0370-2693(88)90916-1}{{\em Phys. Lett.}
  {\bfseries B211} (1988) 363--369}.
%%CITATION = PHLTA,B211,363;%%.

\bibitem{Chivukula:1988gp}
R.~S. Chivukula and A.~V. Manohar, ``{Limits on a Light Higgs Boson},''
  \href{http://dx.doi.org/10.1016/0370-2693(88)90891-X}{{\em Phys. Lett.}
  {\bfseries B207} (1988) 86}.
[Erratum: Phys.~Lett.~B 217, 568 (1989)].
%%CITATION = PHLTA,B207,86;%%.

\bibitem{Patrignani:2016xqp}
{\bfseries Particle Data Group} Collaboration, C.~Patrignani {\em et al.},
  ``{Review of Particle Physics},''
\href{http://dx.doi.org/10.1088/1674-1137/40/10/100001}{{\em Chin. Phys.}
  {\bfseries C40} (2016) 100001}.
%%CITATION = CHPHD,C40,100001;%%.

\bibitem{Leutwyler:1989xj}
H.~Leutwyler and M.~A. Shifman, ``{Light Higgs Particle in Decays of $K$ and
  $\eta$ Mesons},''
\href{http://dx.doi.org/10.1016/0550-3213(90)90475-S}{{\em Nucl. Phys.}
  {\bfseries B343} (1990) 369--397}.
%%CITATION = NUPHA,B343,369;%%.

\bibitem{Kozlov:1995yd}
G.~A. Kozlov, ``{More remarks on search for a new light scalar boson},''
{\em Chin. J. Phys.} {\bfseries 34} (1996) 920--923.
%%CITATION = CJOPA,34,920;%%.

\bibitem{Dawson:1989kr}
S.~Dawson, ``{Higgs Boson Production in Semileptonic $K$ and $\pi$ Decays},''
\href{http://dx.doi.org/10.1016/0370-2693(89)90737-5}{{\em Phys. Lett.}
  {\bfseries B222} (1989) 143--148}.
%%CITATION = PHLTA,B222,143;%%.

\bibitem{Pierog:2013ria}
T.~Pierog, I.~Karpenko, J.~M. Katzy, E.~Yatsenko, and K.~Werner, ``{EPOS LHC:
  Test of collective hadronization with data measured at the CERN Large Hadron
  Collider},'' \href{http://dx.doi.org/10.1103/PhysRevC.92.034906}{{\em Phys.
  Rev.} {\bfseries C92} (2015) 034906},
\href{http://arxiv.org/abs/1306.0121}{{\ttfamily arXiv:1306.0121 [hep-ph]}}.
%%CITATION = ARXIV:1306.0121;%%.

\bibitem{CRMC}
C.~Baus, T.~Pierog, and R.~Ulrich, ``{Cosmic Ray Monte Carlo (CRMC)},''.
  \url{https://web.ikp.kit.edu/rulrich/crmc.html}.

\bibitem{Cacciari:1998it}
M.~Cacciari, M.~Greco, and P.~Nason, ``{The P(T) spectrum in heavy flavor
  hadroproduction},''
  \href{http://dx.doi.org/10.1088/1126-6708/1998/05/007}{{\em JHEP} {\bfseries
  05} (1998) 007},
\href{http://arxiv.org/abs/hep-ph/9803400}{{\ttfamily arXiv:hep-ph/9803400
  [hep-ph]}}.
%%CITATION = HEP-PH/9803400;%%.

\bibitem{Cacciari:2012ny}
M.~Cacciari, S.~Frixione, N.~Houdeau, M.~L. Mangano, P.~Nason, and G.~Ridolfi,
  ``{Theoretical predictions for charm and bottom production at the LHC},''
  \href{http://dx.doi.org/10.1007/JHEP10(2012)137}{{\em JHEP} {\bfseries 10}
  (2012) 137},
\href{http://arxiv.org/abs/1205.6344}{{\ttfamily arXiv:1205.6344 [hep-ph]}}.
%%CITATION = ARXIV:1205.6344;%%.

\bibitem{Cacciari:2015fta}
M.~Cacciari, M.~L. Mangano, and P.~Nason, ``{Gluon PDF constraints from the
  ratio of forward heavy-quark production at the LHC at $\sqrt{s}=7$ and 13
  TeV},'' \href{http://dx.doi.org/10.1140/epjc/s10052-015-3814-x}{{\em Eur.
  Phys. J.} {\bfseries C75} (2015) 610},
\href{http://arxiv.org/abs/1507.06197}{{\ttfamily arXiv:1507.06197 [hep-ph]}}.
%%CITATION = ARXIV:1507.06197;%%.

\bibitem{Kartvelishvili:1977pi}
V.~G. Kartvelishvili, A.~K. Likhoded, and V.~A. Petrov, ``{On the Fragmentation
  Functions of Heavy Quarks Into Hadrons},''
\href{http://dx.doi.org/10.1016/0370-2693(78)90653-6}{{\em Phys. Lett.}
  {\bfseries 78B} (1978) 615--617}.
%%CITATION = PHLTA,78B,615;%%.

\bibitem{Cacciari:2005uk}
M.~Cacciari, P.~Nason, and C.~Oleari, ``{A Study of heavy flavored meson
  fragmentation functions in $e^+ e^-$ annihilation},''
  \href{http://dx.doi.org/10.1088/1126-6708/2006/04/006}{{\em JHEP} {\bfseries
  04} (2006) 006},
\href{http://arxiv.org/abs/hep-ph/0510032}{{\ttfamily arXiv:hep-ph/0510032
  [hep-ph]}}.
%%CITATION = HEP-PH/0510032;%%.

\bibitem{Aaij:2016avz}
{\bfseries LHCb} Collaboration, R.~Aaij {\em et al.}, ``{Measurement of the
  $b$-quark production cross-section in 7 and 13 TeV $pp$ collisions},''
  \href{http://dx.doi.org/10.1103/PhysRevLett.119.169901,
  10.1103/PhysRevLett.118.052002}{{\em Phys. Rev. Lett.} {\bfseries 118} no.~5,
  (2017) 052002}, \href{http://arxiv.org/abs/1612.05140}{{\ttfamily
  arXiv:1612.05140 [hep-ex]}}.
[Erratum: Phys. Rev. Lett.119,no.16,169901(2017)].
%%CITATION = ARXIV:1612.05140;%%.

\bibitem{Bergsma:1985qz}
{\bfseries CHARM} Collaboration, F.~Bergsma {\em et al.}, ``{Search for Axion
  Like Particle Production in 400-{GeV} Proton - Copper Interactions},''
\href{http://dx.doi.org/10.1016/0370-2693(85)90400-9}{{\em Phys. Lett.}
  {\bfseries 157B} (1985) 458--462}.
%%CITATION = PHLTA,157B,458;%%.

\bibitem{Aaij:2015tna}
{\bfseries LHCb} Collaboration, R.~Aaij {\em et al.}, ``{Search for
  hidden-sector bosons in $B^0 \!\to K^{*0}\mu^+\mu^-$ decays},''
  \href{http://dx.doi.org/10.1103/PhysRevLett.115.161802}{{\em Phys. Rev.
  Lett.} {\bfseries 115} (2015) 161802},
\href{http://arxiv.org/abs/1508.04094}{{\ttfamily arXiv:1508.04094 [hep-ex]}}.
%%CITATION = ARXIV:1508.04094;%%.

\bibitem{Aaij:2016qsm}
{\bfseries LHCb} Collaboration, R.~Aaij {\em et al.}, ``{Search for long-lived
  scalar particles in $B^+ \to K^+ \chi (\mu^+\mu^-)$ decays},''
  \href{http://dx.doi.org/10.1103/PhysRevD.95.071101}{{\em Phys. Rev.}
  {\bfseries D95} (2017) 071101},
\href{http://arxiv.org/abs/1612.07818}{{\ttfamily arXiv:1612.07818 [hep-ex]}}.
%%CITATION = ARXIV:1612.07818;%%.

\bibitem{Aad:2013zwa}
{\bfseries ATLAS} Collaboration, G.~Aad {\em et al.}, ``{Characterisation and
  mitigation of beam-induced backgrounds observed in the ATLAS detector during
  the 2011 proton-proton run},''
  \href{http://dx.doi.org/10.1088/1748-0221/8/07/P07004}{{\em JINST} {\bfseries
  8} (2013) P07004},
\href{http://arxiv.org/abs/1303.0223}{{\ttfamily arXiv:1303.0223 [hep-ex]}}.
%%CITATION = ARXIV:1303.0223;%%.

\bibitem{Ferrari:2005zk}
A.~Ferrari, P.~R. Sala, A.~Fasso, and J.~Ranft, ``{FLUKA: A multi-particle
  transport code (Program version 2005)},''.
\url{http://www.fluka.org/fluka.php}.
%%CITATION = CERN-2005-010;%%.

\bibitem{FLUKA2014}
T.~T. Bohlen, F.~Cerutti, M.~P.~W. Chin, A.~Fasso, A.~Ferrari, P.~G. Ortega,
  A.~Mairani, P.~R. Sala, G.~Smirnov, and V.~Vlachoudis, ``{The FLUKA Code:
  Developments and Challenges for High Energy and Medical Applications},'' {\em
  Nuclear Data Sheets} {\bfseries 120} (2014) 211--214.

\bibitem{Mokhov:1995wa}
N.~V. Mokhov, ``{The MARS code system user's guide version 13(95)},''.
  \url{https://mars.fnal.gov}.
FERMILAB-FN-0628.
%%CITATION = FERMILAB-FN-0628;%%.

\bibitem{Mokhov:2012ke}
N.~Mokhov, P.~Aarnio, Y.~Eidelman, K.~Gudima, A.~Konobeev, V.~Pronskikh,
  I.~Rakhno, S.~Striganov, and I.~Tropin, ``{MARS15 Code Developments Driven by
  the Intensity Frontier Needs},''
  \href{http://dx.doi.org/10.15669/pnst.4.496}{{\em Prog. Nucl. Sci. Tech.}
  {\bfseries 4} (2014) 496--501},
\href{http://arxiv.org/abs/1409.0033}{{\ttfamily arXiv:1409.0033
  [physics.acc-ph]}}.
%%CITATION = ARXIV:1409.0033;%%.

\bibitem{Gardner:2015wea}
S.~Gardner, R.~J. Holt, and A.~S. Tadepalli, ``{New Prospects in Fixed Target
  Searches for Dark Forces with the SeaQuest Experiment at Fermilab},''
  \href{http://dx.doi.org/10.1103/PhysRevD.93.115015}{{\em Phys. Rev.}
  {\bfseries D93} (2016) 115015},
\href{http://arxiv.org/abs/1509.00050}{{\ttfamily arXiv:1509.00050 [hep-ph]}}.
%%CITATION = ARXIV:1509.00050;%%.

\bibitem{Bird:2004ts}
C.~Bird, P.~Jackson, R.~V. Kowalewski, and M.~Pospelov, ``{Search for dark
  matter in $b \to s$ transitions with missing energy},''
  \href{http://dx.doi.org/10.1103/PhysRevLett.93.201803}{{\em Phys. Rev. Lett.}
  {\bfseries 93} (2004) 201803},
\href{http://arxiv.org/abs/hep-ph/0401195}{{\ttfamily arXiv:hep-ph/0401195
  [hep-ph]}}.
%%CITATION = HEP-PH/0401195;%%.

\bibitem{Altmannshofer:2009ma}
W.~Altmannshofer, A.~J. Buras, D.~M. Straub, and M.~Wick, ``{New strategies for
  New Physics search in $B \to K^{*} \nu \bar{\nu}$, $B \to K \nu \bar{\nu}$
  and $B \to X_{s} \nu \bar{\nu}$ decays},''
  \href{http://dx.doi.org/10.1088/1126-6708/2009/04/022}{{\em JHEP} {\bfseries
  04} (2009) 022},
\href{http://arxiv.org/abs/0902.0160}{{\ttfamily arXiv:0902.0160 [hep-ph]}}.
%%CITATION = ARXIV:0902.0160;%%.

\bibitem{Khachatryan:2016whc}
{\bfseries CMS} Collaboration, V.~Khachatryan {\em et al.}, ``{Searches for
  invisible decays of the Higgs boson in pp collisions at sqrt(s) = 7, 8, and
  13 TeV},'' \href{http://dx.doi.org/10.1007/JHEP02(2017)135}{{\em JHEP}
  {\bfseries 02} (2017) 135},
\href{http://arxiv.org/abs/1610.09218}{{\ttfamily arXiv:1610.09218 [hep-ex]}}.
%%CITATION = ARXIV:1610.09218;%%.

\bibitem{Aad:2015txa}
{\bfseries ATLAS} Collaboration, G.~Aad {\em et al.}, ``{Search for invisible
  decays of a Higgs boson using vector-boson fusion in $pp$ collisions at
  $\sqrt{s}=8$ TeV with the ATLAS detector},''
  \href{http://dx.doi.org/10.1007/JHEP01(2016)172}{{\em JHEP} {\bfseries 01}
  (2016) 172},
\href{http://arxiv.org/abs/1508.07869}{{\ttfamily arXiv:1508.07869 [hep-ex]}}.
%%CITATION = ARXIV:1508.07869;%%.

\bibitem{Aaboud:2017bja}
{\bfseries ATLAS} Collaboration, M.~Aaboud {\em et al.}, ``{Search for an
  invisibly decaying Higgs boson or dark matter candidates produced in
  association with a $Z$ boson in $pp$ collisions at $\sqrt{s} =$ 13 TeV with
  the ATLAS detector},''
\href{http://arxiv.org/abs/1708.09624}{{\ttfamily arXiv:1708.09624 [hep-ex]}}.
%%CITATION = ARXIV:1708.09624;%%.

\bibitem{Krnjaic:2015mbs}
G.~Krnjaic, ``{Probing Light Thermal Dark-Matter With a Higgs Portal
  Mediator},'' \href{http://dx.doi.org/10.1103/PhysRevD.94.073009}{{\em Phys.
  Rev.} {\bfseries D94} (2016) 073009},
\href{http://arxiv.org/abs/1512.04119}{{\ttfamily arXiv:1512.04119 [hep-ph]}}.
%%CITATION = ARXIV:1512.04119;%%.

\bibitem{Darme:2017glc}
L.~Darmé, S.~Rao, and L.~Roszkowski, ``{Light dark Higgs boson in minimal
  sub-GeV dark matter scenarios},''
\href{http://arxiv.org/abs/1710.08430}{{\ttfamily arXiv:1710.08430 [hep-ph]}}.
%%CITATION = ARXIV:1710.08430;%%.

\bibitem{Angloher:2015ewa}
{\bfseries CRESST} Collaboration, G.~Angloher {\em et al.}, ``{Results on light
  dark matter particles with a low-threshold CRESST-II detector},''
  \href{http://dx.doi.org/10.1140/epjc/s10052-016-3877-3}{{\em Eur. Phys. J.}
  {\bfseries C76} (2016) 25},
\href{http://arxiv.org/abs/1509.01515}{{\ttfamily arXiv:1509.01515
  [astro-ph.CO]}}.
%%CITATION = ARXIV:1509.01515;%%.

\bibitem{Agnese:2015nto}
{\bfseries SuperCDMS} Collaboration, R.~Agnese {\em et al.}, ``{New Results
  from the Search for Low-Mass Weakly Interacting Massive Particles with the
  CDMS Low Ionization Threshold Experiment},''
  \href{http://dx.doi.org/10.1103/PhysRevLett.116.071301}{{\em Phys. Rev.
  Lett.} {\bfseries 116} (2016) 071301},
\href{http://arxiv.org/abs/1509.02448}{{\ttfamily arXiv:1509.02448
  [astro-ph.CO]}}.
%%CITATION = ARXIV:1509.02448;%%.

\bibitem{Akerib:2016vxi}
{\bfseries LUX} Collaboration, D.~S. Akerib {\em et al.}, ``{Results from a
  search for dark matter in the complete LUX exposure},''
  \href{http://dx.doi.org/10.1103/PhysRevLett.118.021303}{{\em Phys. Rev.
  Lett.} {\bfseries 118} (2017) 021303},
\href{http://arxiv.org/abs/1608.07648}{{\ttfamily arXiv:1608.07648
  [astro-ph.CO]}}.
%%CITATION = ARXIV:1608.07648;%%.

\bibitem{Amole:2017dex}
{\bfseries PICO} Collaboration, C.~Amole {\em et al.}, ``{Dark Matter Search
  Results from the PICO-60 C$_3$F$_8$ Bubble Chamber},''
  \href{http://dx.doi.org/10.1103/PhysRevLett.118.251301}{{\em Phys. Rev.
  Lett.} {\bfseries 118} (2017) 251301},
\href{http://arxiv.org/abs/1702.07666}{{\ttfamily arXiv:1702.07666
  [astro-ph.CO]}}.
%%CITATION = ARXIV:1702.07666;%%.

\bibitem{Aprile:2017iyp}
{\bfseries XENON} Collaboration, E.~Aprile {\em et al.}, ``{First Dark Matter
  Search Results from the XENON1T Experiment},''
\href{http://arxiv.org/abs/1705.06655}{{\ttfamily arXiv:1705.06655
  [astro-ph.CO]}}.
%%CITATION = ARXIV:1705.06655;%%.

\bibitem{Cui:2017nnn}
{\bfseries PandaX-II} Collaboration, X.~Cui {\em et al.}, ``{Dark Matter
  Results From 54-Ton-Day Exposure of PandaX-II Experiment},''
\href{http://arxiv.org/abs/1708.06917}{{\ttfamily arXiv:1708.06917
  [astro-ph.CO]}}.
%%CITATION = ARXIV:1708.06917;%%.

\bibitem{Aad:2012tfa}
{\bfseries ATLAS} Collaboration, G.~Aad {\em et al.}, ``{Observation of a new
  particle in the search for the Standard Model Higgs boson with the ATLAS
  detector at the LHC},''
  \href{http://dx.doi.org/10.1016/j.physletb.2012.08.020}{{\em Phys. Lett.}
  {\bfseries B716} (2012) 1--29},
\href{http://arxiv.org/abs/1207.7214}{{\ttfamily arXiv:1207.7214 [hep-ex]}}.
%%CITATION = ARXIV:1207.7214;%%.

\bibitem{Chatrchyan:2012xdj}
{\bfseries CMS} Collaboration, S.~Chatrchyan {\em et al.}, ``{Observation of a
  new boson at a mass of 125 GeV with the CMS experiment at the LHC},''
  \href{http://dx.doi.org/10.1016/j.physletb.2012.08.021}{{\em Phys. Lett.}
  {\bfseries B716} (2012) 30--61},
\href{http://arxiv.org/abs/1207.7235}{{\ttfamily arXiv:1207.7235 [hep-ex]}}.
%%CITATION = ARXIV:1207.7235;%%.

\bibitem{Ade:2015xua}
{\bfseries Planck} Collaboration, P.~A.~R. Ade {\em et al.}, ``{Planck 2015
  results. XIII. Cosmological parameters},''
  \href{http://dx.doi.org/10.1051/0004-6361/201525830}{{\em Astron. Astrophys.}
  {\bfseries 594} (2016) A13},
\href{http://arxiv.org/abs/1502.01589}{{\ttfamily arXiv:1502.01589
  [astro-ph.CO]}}.
%%CITATION = ARXIV:1502.01589;%%.

\bibitem{Abazajian:2016yjj}
{\bfseries CMB-S4} Collaboration, K.~N. Abazajian {\em et al.}, ``{CMB-S4
  Science Book, First Edition},''
\href{http://arxiv.org/abs/1610.02743}{{\ttfamily arXiv:1610.02743
  [astro-ph.CO]}}.
%%CITATION = ARXIV:1610.02743;%%.

\bibitem{Raggi:2015yfk}
M.~Raggi and V.~Kozhuharov, ``{Results and perspectives in dark photon
  physics},''
\href{http://dx.doi.org/10.1393/ncr/i2015-10117-9}{{\em Riv. Nuovo Cim.}
  {\bfseries 38} (2015) 449--505}.
%%CITATION = RNCIB,38,449;%%.

\end{thebibliography}\endgroup

\end{document}